
\def\scalehalf{
  \font\tenrm=cmr10 scaled \magstephalf
  \font\tenbf=cmbx10 scaled \magstephalf
  \font\tenit=cmti10 scaled \magstephalf
  \font\tensl=cmsl10 scaled \magstephalf
  \font\tentt=cmtt10 scaled \magstephalf
  \font\tenex=cmex10 scaled \magstephalf
  \font\teni=cmmi10 scaled \magstephalf
  \font\tensl=cmsl10 scaled \magstephalf
  \font\tensy=cmsy10 scaled \magstephalf
  \font\seveni=cmmi8
  \font\sevenrm=cmr8
  \font\sevensy=cmsy8
  \font\fivei=cmmi6
  \font\fiverm=cmr6
  \font\fivesy=cmsy6
  \font\rm=cmr10 scaled \magstephalf
  \font\bf=cmbx10 scaled \magstephalf
  \font\it=cmti10 scaled \magstephalf
  \font\sl=cmsl10 scaled \magstephalf
  \font\tt=cmtt10 scaled \magstephalf
  \normalbaselineskip 13truept
  \mathfamilydefs}
\def\mathfamilydefs{
  \def\rm{\fam0 \tenrm}
  \def\it{\fam\itfam \tenit}
  \def\sl{\fam\slfam \tensl}
  \def\bf{\fam\bffam \tenbf}
  \def\tt{\fam\ttfam \tentt}
  \def\mit{\fam1 }
  \def\cal{\fam2 }
  \textfont0=\tenrm  \scriptfont0=\sevenrm  \scriptscriptfont0=\fiverm
  \textfont1=\teni  \scriptfont1=\seveni  \scriptscriptfont1=\fivei
  \textfont2=\tensy  \scriptfont2=\sevensy  \scriptscriptfont2=\fivesy
  \textfont\itfam=\tenit
  \textfont\slfam=\tensl
  \textfont\bffam=\tenbf \scriptfont\bffam=\sevenbf
       \scriptscriptfont\bffam=\fivebf
  \textfont\ttfam=\tentt
  \normalbaselines\tenrm}
  
\font\eightrm=cmr8
\font\eighti=cmmi8
\skewchar\eighti='177
\font\eightsy=cmsy8
\skewchar\eightsy='60
\font\eightit=cmti8
\font\eightsl=cmsl8
\font\eightbf=cmbx8
\font\eighttt=cmtt8
\def\eightpoint{\textfont0=\eightrm \scriptfont0=\fiverm
                \def\rm{\fam0\eightrm}\relax
                \textfont1=\eighti \scriptfont1=\fivei
                \def\mit{\fam1}\def\oldstyle{\fam1\eighti}\relax
                \textfont2=\eightsy \scriptfont2=\fivesy
                \def\cal{\fam2}\relax
                \textfont3=\tenex \scriptfont3=\tenex
                \def\it{\fam\itfam\eightit}\relax
                \textfont\itfam=\eightit
                \def\sl{\fam\slfam\eightsl}\relax
                \textfont\slfam=\eightsl
                \def\bf{\fam\bffam\eightbf}\relax
                \textfont\bffam=\eightbf \scriptfont\bffam=\fivebf
                \def\tt{\fam\ttfam\eighttt}\relax
                \textfont\ttfam=\eighttt
                \setbox\strutbox=\hbox{\vrule
                     height7pt depth2pt width0pt}\baselineskip=10pt
                \rm}
\font \bigone=cmbx10 scaled\magstep1

\def\d3k{{\displaystyle d{\bf k} \over \displaystyle (2\pi)^3}}

\def\Del{{\rm Del}}
\def\Vor{{\rm Vor}}
\def\mg{\big <}
\def\md{\big >}
\def\hang{\par\hangindent=\parindent\@}
\def\@{\noindent}

\def\lessapprox{\,\raise 0.6ex\hbox{$<$}\kern -0.75em\lower 0.47ex
    \hbox{$\sim$}\,}
\def\largapprox{\,\raise 0.6ex\hbox{$>$}\kern -0.75em\lower 0.47ex
    \hbox{$\sim$}\,}
%
%
\newcount\eqnumber
\eqnumber=1
\def\step#1{\global\advance#1 by 1}
\def\neweq{{\rm\the\eqnumber}\step{\eqnumber}}
\def\eqnew{\eqno(\neweq)}
{\scalehalf
\newcount\startpage
\startpage=1
\pageno=\startpage
\baselineskip 16pt
\voffset=-0.5truecm
\overfullrule=0pt
\tolerance=1500
\centerline{\bf{\bigone A NEW METHOD FOR ACCURATE VELOCITY STATISTICS
ESTIMATION}}
\bigskip
\centerline{\bf Francis Bernardeau $^{1,4}$ \& Rien van de Weygaert $^{2,3,4}$}
\medskip
\vbox{\noindent\eightpoint
\centerline{$^{1}$ Service de Physique Th\'eorique, CE de Saclay,
F-91191 Gif-sur-Yvette, Cedex France}
\centerline{$^{2}$ Max-Planck-Institut f\"ur Astrophysik,
Karl-Schwarzschild-Stra{\ss}e 1, Garching bei M\"unchen, Germany}
\centerline{$^{3}$ Kapteyn Astronomical Institute, University
of Groningen, P.O. Box 800, 9700 AV Groningen, the Netherlands}
\centerline{$^{4}$ Canadian Institute for Theoretical Astrophysics,
60 St. George Street, Toronto, Ontario M5S 1A7, Canada}
\smallskip
\centerline{Email: fbernard@amoco.saclay.cea.fr, weygaert@astro.rug.nl}
}
\bigskip
\centerline{\bf ABSTRACT}
\medskip
\vbox{\baselineskip 12pt
We introduce two new methods to obtain reliable velocity
field statistics from N-body simulations, or indeed from any general
density and velocity fluctuation field sampled by discrete points. These
methods, the {\it Voronoi tessellation method} and {\it Delaunay
tessellation method}, are based on the use of the Voronoi and
Delaunay tessellations of the point distribution defined by the
locations at which the velocity field is sampled. In the Voronoi
method the velocity is supposed to be uniform within the Voronoi
polyhedra, whereas the Delaunay method constructs a velocity field by
linear interpolation between the four velocities at the locations
defining each Delaunay tetrahedron.

The most important advantage of these methods is that they provide
an optimal estimator for determining the statistics of volume-averaged
quantities, as opposed to the available numerical methods that
mainly concern mass-averaged quantities. As the major share of
the related analytical work on velocity field statistics has focussed
on volume-averaged quantities, the availability of appropriate
numerical estimators is of crucial importance for checking the
validity of the analytical perturbation calculations. In addition,
it allows us to study the statistics of the velocity field in the
highly nonlinear clustering regime.

Specifically we describe in this paper how to estimate, in both the
Voronoi and the Delaunay method, the value of the volume-averaged
expansion scalar $\theta \equiv H^{-1} {\bf \nabla \cdot v}$ --  the
divergence of the peculiar velocity, expressed in units of the Hubble
constant $H$, as well as the
value of the shear and the vorticity of the velocity field, at an
arbitrary position. The evaluation of these quantities on a regular
grid leads to an optimal estimator for determining the
probability distribution function (PDF) of the volume-averaged
expansion scalar, shear and vorticity.
Although in most cases both the Voronoi and the Delaunay method
lead to equally good velocity field estimates, the Delaunay method may
be slightly preferable. In particular it performs considerably better
at small radii. Note that it is more CPU-time intensive while its
requirement for memory space is almost a factor 8 lower than the
Voronoi method.

As a test we here apply our estimator on an N-body
simulation of such structure formation scenarios. The PDFs determined
from the simulations agree very well with the analytical predictions.
An important benefit of the present method is that, unlike previous
methods, it is capable of probing accurately the velocity field
statistics in regions of very low density, which in N-body simulations
are typically sparsely sampled.

In a forthcoming paper we will apply the newly developed tool to
a plethora of structure formation scenarios, both of Gaussian and
non-Gaussian initial conditions, in order to see in how far the
velocity field PDFs are sensitive discriminators, highlighting
fundamental physical differences between the scenarios.}

\medskip
\@\medskip
\@{\it Subject headings:} Cosmology: theory -- large-scale structure
of the Universe -- Methods: numerical -- statistical

\medskip
\@{\bf{\bigone 1. Introduction}}
\smallskip
\@Besides the distribution of galaxies and the temperature fluctuations in the
cosmic microwave background, and possibly weak gravitational lensing of
background galaxies by large scale structures (see e.g. Villumsen
1995), the velocity field on
cosmological scales  is one of the main sources of information on the
formation and evolution
of structure in the Universe. Early work indicated the existence of
large-scale velocity flows (Rubin et al. 1976) and established
the existence of the motion of the Local Group with respect
to the rest-frame defined by the microwave background (Smoot \& Lubin
1979). However, it was the work by Burstein et al. (1987) that established
beyond doubt that the Local Group is participating in a large scale
streaming motion.

The advent of reliable redshift-independent distance estimators lead
to an enormous growth of activity in the field of measuring and
interpreting the peculiar velocities of galaxies. This growth of
attention was evidently fed by the fact that the velocity field
provides direct information on the dynamics of the Universe on scales
of more than a few $\rm Mpc$. Above these scales dynamical relaxation
has not had yet a chance to wash out the memory of the
conditions in the early Universe. The velocity field can therefore be
fruitfully investigated by means of perturbation analysis.
Particularly useful is the, $\Omega$-dependent, velocity-density
relationship that follows from linear theory (see e.g. Peebles
1980). Moreover, a general result of perturbation theory
predicts that the rotational part of the
velocity field vanishes. Based on this observation Bertschinger \&
Dekel (1989) developed the non-parametric
POTENT method in which the
local cosmological velocity field is reconstructed from the measured
line-of-sight velocities. In a series of papers (e.g. Bertschinger et al.
1990), they applied their method to the
existing catalogues of galaxy peculiar velocities.

The POTENT analysis also paves the way for a more quantitative analysis of the
velocity field, estimating various statistical properties and their
relation to the properties of the density field. For instance, via the
velocity-density relationship it is possible to reconstruct the
corresponding density field or, by comparison with a uniformly sampled
galaxy redshift catalogue (in particular the IRAS based redshift
catalogues, e.g. Strauss et al. 1990), obtain an idea of in how far
the galaxy distribution forms a biased tracer of the underlying mass
distribution. If it is assumed that this bias can be simply
represented by a linear bias factor $b$, the value of $\Omega^{0.6}/b$
can be estimated from the measured peculiar velocity field and
a uniformly sampled redshift catalogue (e.g. Yahil et al. 1991).

A variety of other methods have been proposed to determine this
combination of $\Omega$ and $b$ (see the review paper of Dekel 1994 and
references therein). On the basis of a more specific statistical
analysis, in particular of the velocity divergence, other studies
managed to determine these two parameters separately. For example,
Nusser \& Dekel (1993) proposed to use a reconstruction method assuming
Gaussian initial conditions to constrain $\Omega$, while Dekel
\& Rees (1994) used voids to achieve the same goal.

Another approach has been proposed by Bernardeau et al. (1995) and
Bernardeau (1994a). This is based on the use of statistical properties
of the divergence of the locally smoothed velocity field, its skewness
(the $3^{rd}$ order moment) and its kurtosis ($4^{th}$ order moment).
It was shown that these statistical quantities are potentially
very valuable to measure $\Omega$ or to test the gravitational
instability scenario. The early analytical work was subsequently
extended towards the determination of the complete velocity divergence
probability distribution function for gravitational instability scenarios
starting from Gaussian initial conditions in the case the velocity
field is filtered by a top-hat window function. Preliminary comparisons
with numerical simulations (Bernardeau 1994b, Juszkiewicz et al. 1995,
\L okas et al. 1995) yielded encouraging results. However, a comparison of
the analytical
results with the PDFs determined from N-body simulations is
complicated considerably by the fact that in the latter velocities
are only known at, non-uniformly distributed, particle locations.

In this paper we address specifically the issue of the discrete nature
of the velocity sampling, which leads to the development of a
numerical method to obtain reliable velocity field statistics from
N-body simulations. The goal is threefold. First of all,
we wish to have an independent way of checking whether the
perturbation calculations that yield the quasi-linear results are
indeed valid. Secondly, if so, these analytical results form a good
`calibration' point for the numerical tool that we have developed
here, so that we can use it with confidence in highly nonlinear
conditions. Finally, because the velocity field in observational
samples is also only known at a discrete number of positions,
the location of galaxies, we may be able to apply the
developed method, in adapted form, to the available catalogues of
measured galaxy peculiar velocities.

The central problem that we address here is that while almost all analytical
results concern volume-averaged quantities, almost all available
numerical estimators in essence only yield mass-averaged
quantities. This may considerably complicate
any comparison, and even lead to false conclusions
regarding e.g. the validity of perturbation theory. To improve upon this
situation we introduce two new numerical methods, the {\it Voronoi
tessellation method} and the {\it Delaunay tessellation method}.
Both methods are based on two important objects in stochastic
geometry, the Voronoi and the Delaunay tessellation of the point set
consisting of the points at which the velocity field has been
sampled. A Voronoi tessellation of a set of nuclei is a space-filling
network of polyhedral cells, each of which delimit that part of space
containing that is closer to its nucleus than to any of
the other nuclei. The Delaunay tessellation is also a space-filling
network of mutual disjunct objects, tetrahedra in 3-D. The four
vertices of each Delaunay tetrahedron are nuclei from the point set,
such that the corresponding circumscribing sphere does not
have any of the other nuclei inside. The Voronoi and the Delaunay
tessellation are closely related, and are dual in the sense that one
can be obtained from the other.

The earliest use of Voronoi tessellations can be found in the work of
Dirichlet (1850) and Voronoi (1908) in their work on the reducibility of
positive definite quadratic forms. However, their application to random
point patterns has caused this concept to arise independently in various
fields of science and technology, ranging from molecular physics to
forestry (see Stoyan, Kendall, and Mecke 1987 for references). Because of
these diverse applications they acquired a set of alternative names, such as
Dirichlet regions, Wigner-Seitz cells, and Thiessen figures. Despite
the simplicity of its definition, analytical work on the statistics
of Voronoi tessellations has appeared to be rather complicated and cumbersome,
so that present analytical knowledge is mainly restricted to a few
statistical moments of the distribution function of geometrical
properties of Voronoi
tessellations generated by homogeneous Poisson processes (see e.g. Meyering
1953, Gilbert 1962, Miles 1970, and M{\o}ller 1989). For the time
being, numerical work is therefore an inescapable necessity for any
progress in this field (see e.g. Van de Weygaert 1991b, 1994).

Within astronomy, and in particular cosmology, Voronoi tessellations are
mostly known for their application as geometrical models for astrophysical
structures. One of the first applications was by Kiang (1966), who used
them in an attempt to derive a mass spectrum resulting from the fragmentation
of interstellar clouds. The similarity of Voronoi tessellations to the
cellular patterns in the galaxy distribution sparked a lot of work in
cosmology. Matsuda \& Shima (1984) pointed out the similarity between
two-dimensional Voronoi
tessellations and the outcome of numerical clustering simulations in a
neutrino-dominated universe. Independently,
Voronoi tessellations were introduced into cosmology in a study by
Icke \& Van de Weygaert (1987) of the statistical properties of
two-dimensional Voronoi tessellations. They argued that a cellular pattern
similar to Voronoi tessellations is a natural outcome of the evolution
of a void-dominated Universe. Their statistical analysis was extended to three
dimensions in Van de Weygaert (1991b, 1994), based on the completion of a
three-dimensional geometrical Voronoi algorithm. In an early analysis of
three-dimensional Voronoi tessellations, Van de
Weygaert \& Icke (1989) found that Voronoi tessellations also possess
some interesting clustering properties, as was confirmed in a
Monte-Carlo study by Yoshioka \& Ikeuchi (1989). Since then, the amount
of applications of Voronoi tessellations as a useful, conceptually simple
model for a cellular or foam-like distribution of galaxies on large scales,
has steadily increased (Coles 1990; Van de Weygaert 1991a,b;
Ikeuchi \& Turner 1991; SubbaRao \& Szalay 1992; Williams 1992;
Williams et al. 1992; Goldwirth, Da Costa, \& Van de Weygaert 1995).

However, their use as a geometrical model is a different class
of applications of Voronoi tessellations than the one we use in the
present paper. Here we follow another philosophy, namely that the
sensitivity of the geometrical characteristics of the Voronoi
tessellations to the underlying nucleus
distribution makes the Voronoi tessellation, and its dual the Delaunay
tessellation, a potentially very useful instrument to study the
properties of a point process. Their usefulness as statistical descriptors
was suggested earlier by e.g. Finney (1979), who introduced the name
`polyhedral statistics'. Instead of being interested in the generating
point process itself, we turn our attention to developing a reliable
and/or optimal description of the velocity field sampled by the point
process.

The first method that we have introduced here is based on the Voronoi
tessellations, and follows
directly from the definition of volume-averaged velocities (eqn. 3).
It can be considered as the multidimensional extension of the
approximation of a function of one variable by a constant value in
a finite number of bins, the constant value being equal to the function value
at the point in the bin. The {\it Voronoi method} yields a
velocity field with constant values of the velocity components within
each Voronoi cell of the defining point distribution. The velocity
within the whole interior of the cell is equal to the velocity
of the nucleus of that cell. Consequently, only at the boundaries of
the Voronoi cells the velocity gradient has a non-zero value. The
subsequent operation of volume-averaging of a quantity therefore consists of
determining the intersection of Voronoi walls with the appropriate
filter, for a top-hat filter a sphere of radius $R$, and weighing the
value of that quantity in the wall with the size of intersection area.

The {\it Delaunay method}, on the other hand, should be regarded as the
multidimensional recipe for linear interpolation. In a space of
dimension $d \geq 2$ linear interpolation consists of assuming
constant function gradients in interpolation intervals defined by
$d+1$ points,
`hyper-triangles'. In two-dimensional space a hyper-triangle is a
triangle, in three-dimensional space a tetrahedron. Unlike the
one-dimensional case, the choice of multidimensional interpolation
tetrahedra may not be unique. However, here we argue that Delaunay
tetrahedra are a natural and logical choice based on the requirements
that the whole of the sample space is
covered by a space-filling network of mutually disjunct tetrahedra and
that these tetrahedra should be as compact as possible to minimize
approximation errors. The (constant) value of each of the
9 velocity field gradients in each of the Delaunay tetrahedra is
determined from the locations of and velocities at the four points
that define each of them. Summarizing, the Delaunay method consists of
three major steps: (1) construction of the Delaunay tessellation,
(2) determination of the 9 velocity gradients $\partial v_i / \partial
x_j$ in each of the tetrahedra, (3) volume averaging of the obtained
field of velocity gradients. For a top-hat filter the latter step
consists of determining the volume of the intersection of
the Delaunay tetrahedra with the filter sphere.

In this paper, we start by shortly discussing the conventional
methods to sample the value of velocity fields on grid positions from
the value of the velocity at a discrete number of points. In particular
we stress the fundamental difference between mass and volume weighted
velocity averages. Conventional estimators are almost always based on
mass weighted velocity fields. This may induce
complications in the comparison with theoretical predictions. This
leads to the introduction in section 3 of the Voronoi and the Delaunay
method. Both are good estimators for volume weighted velocity fields, and are
therefore instrumental in improving comparisons between theoretical
predictions and the results of N-body simulations. The accompanying
appendices A and B contain detailed
descriptions of the geometrical details of these sampling techniques.
Computational considerations are discussed in section 4, while section
5 contains a discussion of an application of both methods to an N-body
simulation of galaxy clustering, determining the values of the
divergence, vorticity and shear of the velocity field on a regular
grid. These values are
compared with each other as well as with the ones determined by a
conventional estimator. Finally, in section 6 we conclude with a discussion
of the virtues of the new methods, and suggestions for future applications.

\medskip
\@{\bf {\bigone 2. Discretely sampled velocity field}}
\smallskip
The fact that the velocity field is only known at a finite number
of discrete positions is a major technical obstacle for obtaining
reliable estimates of statistical parameters of the velocity field.
It is of importance both in the observational data as well as in N-body
simulations. One possibility is to smooth the galaxy velocity field
with a filter. For example, Bertschinger et al. (1990) choose to
filter the measured galaxy velocities with a Gaussian smoothing
function. In one case they took a fixed smoothing length, to be
preferred for a rigorous statistical analysis, while for an
optimal representation of the velocity and density field they
adopted a filter with an adaptive smoothing length. By taking
this length to be equal to the distance to the fourth nearest
neighbour the analysis automatically becomes better in the
well-sampled high-density regions, thereby reducing the problem of
noisy data and sparsely sampled underdense regions. In their analysis of
numerical simulations Juszkiewicz et al. (1995) and {\L}okas et al. (1995)
also used smoothing of the velocity field by a Gaussian filter
with a fixed smoothing length to obtain the local velocity field
on a regular grid.

Effectively these are mass weighted velocity fields,
$${\bf v}_{mass}({\bf x}_0) \equiv {\displaystyle \int
d{\bf x}\,{\bf v}({\bf x}) \,\rho({\bf x}) W_M({\bf x},{\bf x}_0) \over
\displaystyle \int d{\bf x}\,\rho({\bf x}) W_M({\bf x},{\bf x}_0)}
\,,\eqnew$$
\@where $W_M({\bf x},{\bf x}_0)$ is the used filter function defining
the weight of a mass element in a way dependent on its position
relative with respect to the position ${\bf x}_0$. In other words,
${\bf v}_{mass}$ is effectively the velocity corresponding to the
average momentum within the filter volume. For a discrete particle
distribution ${\bf v}_{mass}$ reduces to,
$$\eqalign{{\bf v}_{mass}({\bf x}_0)\,&=\,{\displaystyle \int d{\bf x}\,{\bf
v}({\bf x})\,
W_M({\bf x},{\bf x}_0)\,\sum_i \delta_D({\bf x}-{\bf x}_i) \over \displaystyle
\int d{\bf x}\,W_M({\bf x},{\bf x}_0) \sum_i \delta_D({\bf x}-{\bf x}_i)}\cr
 &=\, {\displaystyle \sum_i w_i {\bf v}({\bf x}_i) \over
\displaystyle \sum_i w_i}\,.\cr}\eqnew$$
\@where $w_i \equiv W_M({\bf x}_i,{\bf x}_0)$ and $\delta_D({\bf
x},{\bf x}_0)$ the Dirac delta function.
A major complication of such an analysis is that a comparison of
statistical properties of the velocity field obtained in this way with
the known analytical results is not straightforward. Almost without
exception these analytical results are based on the volume weighted
filtered velocity field ${\tilde {\bf v}}$,
$${\tilde {\bf v}}({\bf x}_0) \equiv {\displaystyle
\,\int d{\bf x}\,{\bf v}({\bf x}) W_V({\bf x},{\bf x}_0) \over
\displaystyle \int d{\bf x}\,W_V({\bf x},{\bf x}_0)}\,,\eqnew$$
\@where $W_V({\bf x},{\bf x}_0)$ is a possibly used weight function.
At the present the only analytical work that has treated certain
aspects of the statistics of the mass averaged velocity, namely the
skewness of $\theta=H^{-1} \nabla \cdot {\bf v}$, is the work by Bernardeau
et al.  (1995). A major complication is that this quantity involves
the density field, and therefore would introduce the unknown
bias between mass and galaxy density field in a practical
implementation.

In order to get an estimate of the volume averaged velocity
Juszkiewicz et al. (1995) use a two-step scheme, wherein they first
determine a velocity on a grid according to equation (2), and then
use the resulting grid of velocities to determine volume averaged
velocities according to equation (3), the volume averaging being
accomplished by Gaussian filtering. Bernardeau (1994b) followed a similar
procedure, whereby he used a top-hat filter for the volume averaging.
Such a scheme would yield
reliable results if the filter length of $W_M$ would be much smaller
than that of $W_V$. However, for technical reasons this is often
difficult to attain. For example, it requires a very small grid size which
would be excessively computer time and memory consuming.

In the study of {\L}okas et al. (1995) the same approach was followed
in a comparison between N-body simulations and analytical perturbation
calculations. While they found good agreement with the obtained
density field moments, there already appeared to be substantial discrepancies
in the case of the velocity divergence $\theta$ by the
time $\sigma_{\theta} \approx 0.1$. They suggested as possible
causes for the disagreement that (1) the perturbation approximation
already starts to break down by the time $\sigma_{\theta} \approx
0.1$, (2) the N-body simulations do not evolve the velocity divergence
field with sufficient accuracy and (3) the estimator of the smoothed
velocity divergence from the N-body simulations is too noisy or too
inaccurate. In order to improve upon this situation it is therefore
highly desirable to develop a more reliable estimator,
preferentially closer related to the `volume-averaged' nature of the
perturbation calculations. In the following sections we will introduce
two new methods that overcome the dilemma of a required very small
initial smoothing length by constructing the Voronoi and Delaunay
tessellations of the point distribution.

\topinsert
\vbox{\noindent\eightpoint {\bf Figure 1.}
The Delaunay triangulation
(left frame) and Voronoi tessellation (right frame) of a distribution
of 25 nuclei (stars) in a square (central panel). Periodic boundary conditions
are assumed.}
\endinsert

\medskip
\@{\bf {\bigone 3. Voronoi and Delaunay Tessellations}}
\smallskip
The first new estimator that we introduce here is based on Voronoi
tessellations. It follows in a rather direct way from an asymptotic
interpretation of the definition of mass filtered quantities (eqn. 2).
Although it leads to good results, it corresponds to an artificial situation
of a discontinuous velocity field. This is successfully improved upon by
a subsequent further elaboration and extension of the Voronoi method
to a second estimator, based on the division of space into Delaunay
tetrahedrons.

\medskip
\@{\bf{3.1 The Voronoi Tessellation}}
\smallskip
\@In section 2 we made the observation that a good approximation of volume
averaged quantities is obtained by volume averaging over quantities
that were mass filtered with, in comparison, a very small scale for the
mass weighting filter function. We can then make the observation that
the asymptotic limit of this, namely using a filter with an
infinitely small filter length (see eqn. 2),
$$\eqalign{{\bf v}_{mass}({\bf x}_0)\,&=\,{\displaystyle \sum_i w_i {\bf
v}({\bf x}_i)
\over \displaystyle \sum_i w_i}\,\cr
&=\, {{\displaystyle {\bf v}({\bf x_1})+
\sum_{i=2}^N {w_i \over w_1} {\bf v}({\bf x}_i}) \over \displaystyle
1 + \sum_{i=2}^N {w_i \over w_1}}\quad \longrightarrow \quad {\bf v}({\bf
x}_1)\,,\cr}\eqnew$$
\@where we have ordered the locations $i$ by increasing distance to
${\bf x}_0$ and thus by decreasing value of $w_i$, is in fact taking the
velocity ${\bf v}({\bf x}_1)$ of the
closest particle as the estimate of the mass averaged velocity. In
other words, we can divide up space into regions consisting of that
part of space closer to a particle than to any other particle, and
taking the velocity of that particle as the value of the velocity
field in that region. The search for the closest particle of each point of
the field naturally leads to the construction of the Voronoi tessellation
associated with the particle distribution. This division of space is a
familiar and important concept in the field of stochastic geometry
(see Stoyan, Kendall and Mecke 1987 for an overview of this field).
It consists of space-covering and mutual disjunct set of convex cells,
each of which contains
a particle of the original particle distribution, enclosing the points of
space for which the closest particle is precisely the one in the cell.

Formally a Voronoi tessellation (Voronoi 1908, Dirichlet 1850) can be defined
as follows. Assume that
we have a distribution of a countable set $\Phi$ of nuclei $\{x_i\}$ in
$\Re^d$. Let ${\vec x}_1, {\vec x}_2, {\vec x}_3,\ldots$ be the
coordinates of the nuclei. Then, the {\it Voronoi region} $\Pi_i$ of
nucleus $i$ is defined
by the following set of points ${\vec x}$ of the space:
$$\Pi_i = \{{\vec x} \vert d({\vec x},{\vec x}_i) < d({\vec x},{\vec x}_j)
\qquad \hbox{for all $j \not= i$} \},\eqnew$$
\@where $d({\vec x},{\vec y})$ is the Euclidian distance between ${\vec x}$
and ${\vec y}$. In other words, $\Pi_i$ is the set of points which is nearer
to ${\vec x}_i$ than to ${\vec x}_j,\quad j \not=i$. Each region
$\Pi_i$ therefore consists of the intersection of the open half-spaces bounded
by the
perpendicular bisectors of the segments joining ${\vec x}_i$ with each
of the other ${\vec x}_j$'s. Hence, Voronoi regions are convex polyhedra
(3-D) with finite size according to definition (5). Each $\Pi_i$ is called
a {\it Voronoi polyhedron}. The complete set of $\{\Pi_i\}$ constitute
a tessellation of $\Re^d$, the {\it Voronoi tessellation} ${\cal V}(\Phi)$
relative to $\Phi$. A two-dimensional Voronoi tessellation of 25 cells
is shown in figure 1 (right-hand frame), while an idea of a
three-dimensional Voronoi tessellation can be obtained from the three
Voronoi cells displayed in figure 2. The latter three cells are taken
from a Voronoi network of 1000 cells, generated by Poissonian
distributed nuclei, in a box with periodic boundary conditions.

As described above, given a field of velocities at a set of discrete
particles a reasonable first assumption is that the velocity is
constant within each Voronoi cell. The Voronoi method can therefore be
regarded as the three- (or two-) dimensional equivalent of approximating a
one-dimensional function $f(x)$ by a sequence of intervals wherein the
function has a constant value. If the value of the function $f(x)$
is known at $M$ discrete points $x_i$ (upper panel figure 4), this
one-dimensional equivalent of the Voronoi method consists of adopting
a constant value $f(x_i)$ in the interval between $(x_i+x_{i-1})/2$
and $(x_i+x_{i+1})/2$ (see central panel of figure 4).

\topinsert
\vbox{\noindent\eightpoint
{\bf Figure 2.}
Three examples of
three-dimensional Voronoi cells. They are taken from a Voronoi
tessellation generated by 1000 Poissonian distributed nuclei in a
box with periodic boundary conditions. The stars in each of the cells
represent the position of the cell nucleus.}
\endinsert

The first step of the Voronoi algorithm consists of calculating the
Voronoi tessellation that is defined by the set of points at which
the velocity field has been sampled. For this we use the three-dimensional
geometrical Voronoi code that was developed by Van de Weygaert (1991b, 1994).
Starting from an input of points this code calculates the complete
geometrical structure, i.e. the location of the walls, edges and vertices,
of the corresponding Voronoi tessellation.

Subsequently, we proceed by determining the corresponding volume averaged
quantities. This is accomplished by a volume filtering of the resulting
velocity field. By adopting a top-hat filter $W_{TH}$ with radius $R$ as the
volume filter $W_V$, the problem of determining the corresponding
volume-averaged velocity gradient ${\tilde v_{i,j}} \equiv
{\widetilde {\nabla_j v_i}}$ is determining the average value of $v_{i,j}$
within a sphere of radius $R$,
$$\eqalign{{\tilde v_{i,j}}({\bf x}_0) \,= \, {\widetilde {\displaystyle
\partial v_i
\over \partial x_j}} \,&\equiv \,{\displaystyle \,\int d{\bf x}\,
v_{i,j}({\bf x}) W_{TH}({x},{\bf x}_0) \over \displaystyle
\int d{\bf x}\,W_{TH}({\bf x},{\bf x}_0)}
\,\cr
&=\,{\displaystyle 3 \over \displaystyle 4 \pi R^3}\,
\int_R d{\bf x}\, v_{i,j}\,,\cr}\eqnew$$
\@where the latter volume integral is over the part of space enclosed by the
sphere with radius $R$ centered on ${\bf x}_0$. The constant value of the
velocity ${\bf v}$ within each Voronoi cell automatically implies that the
value of the velocity gradient $v_{i,j} ({\bf x})$ is
equal to zero in their interior, so that the cells themselves have a
contribution zero to the integral $\int d{\bf x} \, v_{i,j}$. Only at the
boundaries between the Voronoi cells $v_{i,j}$ will have a non-zero value.
Moreover, the value of $v_{i,j}$ will be constant within each wall, as it
corresponds to the change of value of ${\bf v}$ between the two corresponding
neighbour cells (see figure 3). The finite contribution by
any Voronoi wall that lies within or intersects the filter sphere can
be calculated by considering the volume defined by the surface of the
wall and having an infinitesimal width $\Delta s$ perpendicular to
the wall. Imagine a Voronoi wall $k$,
being the boundary between two Voronoi cells $k1$ and $k2$ in whose interior
the velocities are ${\bf v}_{k1}$ and ${\bf v}_{k2}$ (see figure 3),
that intersects the top-hat sphere.
The wall has a perpendicular width $\Delta s_k$ and a surface area
$A_k$ within the sphere,
while its orientation is determined by its normal vector ${\bf n}_k$.
The velocity gradient $v_{x,y}$ can then be easily inferred from the velocity
change $\Delta v_x = ({\bf v}_{k2}-{\bf v}_{k1}) \cdot {\bf e}_x$ along the
$x$-direction. Because this corresponds to an interval $\Delta y =
({\bf n}_k \cdot {\bf e}_y)\,\Delta s_k$ in the $y$ direction, we find
$${\displaystyle \partial v_x \over \displaystyle \partial y} \,\approx\,
{\displaystyle \Delta v_x \over \displaystyle \Delta y} \,=\,
{\displaystyle ({\bf n}_k \cdot {\bf e}_y) \ ({\bf v}_{k2}-{\bf v}_{k1})
\cdot {\bf e}_x \over \displaystyle \Delta s_k}\,.\eqnew$$
\@This can be generalized directly to yield the following result for the
volume averaged value ${\tilde v_{i,j}}$ (with $(i,j)=1,2,3$),
$${\tilde v_{i,j}}\,=\, {\displaystyle 3 \over \displaystyle 4 \pi
R^3}\,\sum_k\,A_k\,({\bf n}_k \cdot {\bf e}_j)
\ ({\bf v}_{k1}-{\bf v}_{k2})\cdot {\bf e}_j
\,,\eqnew$$
\@where the sum is over all walls that intersect the sphere.
In the specific case of the volume-averaged velocity divergence
${\tilde \theta}$, this leads to the expression
$$\tilde \theta \,=\,{\displaystyle \nabla \cdot {\tilde {\bf v}}
\over \displaystyle H} = {\displaystyle
3 \over \displaystyle 4 \pi R^3}\,\sum_k\,{\displaystyle A_k\,
{\bf n}_k\cdot({\bf v}_{k1}-{\bf v}_{k2})\over \displaystyle H}\,.\eqnew$$
\@The problem of determining the volume-averaged velocity velocity gradients
(eqn. 8), or velocity divergence (eqn. 9), has therefore been reduced to
determining the intersection of the walls in the Voronoi tessellations with
spheres. This is a geometrical problem that can be solved with some effort
(see Appendix A).

\topinsert
\vbox{\noindent\eightpoint
{\bf Figure 3.}
Two-dimensional illustration
of the Voronoi method to estimate velocity field gradients (compare
eqn. 7-9). The stars indicate the positions of the nuclei of the
Voronoi cells. Part of the cell boundaries are indicated by the walls
with finite thickness. The dashed arrows at each of the
stars are the velocity vectors at each of these locations. The solid
arrow is the normal vector ${\vec n}_k$ of the wall $k$, of thickness
${\Delta s}_k$, between the two cells $k_1$ and $k_2$, in whose interior
the velocity field has values ${\vec v}_{k1}$ and ${\vec v}_{k2}$
respectively. Partially indicated is the circle corresponding to
a top-hat filter centered on a grid position.}
\endinsert

In practice we repeat this procedure of top-hat averaging at each point of
a regular grid. From the values of $\partial v_i / \partial x_j$
at each grid point we can then easily evaluate the value of the velocity
divergence $\theta$, the shear $\sigma_{ij}$ and the vorticity
${\bf \omega}$, where ${\bf \omega}=\nabla \times {\bf v}=
\epsilon^{kij} \omega_{ij}$ (and $\epsilon^{kij}$ is the completely
antisymmetric tensor),
$$\eqalign{\theta\,&=\,
{1 \over H}\left({\displaystyle \partial v_x \over \displaystyle \partial x} +
{\displaystyle \partial v_y \over \displaystyle \partial y} +
{\displaystyle \partial v_z \over \displaystyle \partial z}\right)\,,\cr
\sigma_{ij}\,&=\,
{1 \over 2}\left\{
{\displaystyle \partial v_i \over \displaystyle \partial x_j} +
{\displaystyle \partial v_j \over \displaystyle \partial x_i}
\right\}\,-\, {1 \over 3} \,(\nabla\cdot{\bf v})\,\delta_{ij} \,,\cr
\omega_{ij}\,&=\,
{1 \over 2}\left\{
{\displaystyle \partial v_i \over \displaystyle \partial x_j} -
{\displaystyle \partial v_j \over \displaystyle \partial x_i}
\right\}\,.\cr}\eqnew$$
\@Summarizing, the end result of the operations described above consists
of fields of top-hat averaged quantities like velocity divergence, shear and
vorticity, sampled at regular grid intervals.

\medskip
\@{\bf{3.2 The Delaunay Tessellation}}
\smallskip
A major characteristic of the Voronoi approach is that it leads to a
discontinuous velocity field. This is evidently not the only unique
way of defining the velocity field from a discrete set of sample
points. Moreover, the Voronoi method may have problems in the case
of small filter radii. Relevant non-zero values for the velocity gradients are
only produced in the Voronoi walls. Many filter spheres would remain
empty when the filter scale is smaller than the average distance
between Voronoi walls, i.e. when their scales are smaller than
$\approx L/N^{1/3}$ (with $L$ the boxsize and $N$ the number of
Voronoi cells). This would yield many irrelevant and noisy filter
velocity gradient averages.
It may therefore be worthwhile to define another independent method
based on a different interpolation scheme. In fact, this would give us
the possibility of internally checking the results of the new methods
that we have introduced here.

\topinsert
\vbox{\noindent\eightpoint
{\bf Figure 4.}
Illustration of the
one-dimensional equivalents of the Voronoi and the Delaunay method.
The value of some function $f(x)$ has been determined at 20 randomly
distributed positions $x_i$ along a line (top panel). The height of
a star at position $x_i$ is proportional to the value $f(x_i)$. Central
panel: `Voronoi method'. The function $f$ in the whole bin centered on
a point $x_i$, bounded by the points halfway in between $x_i$ and its
neighbours $x_{i-1}$ and $x_{i+1}$, is assumed to have the constant
value $f(x_i)$. Note that these bins are essentially one-dimensional
Voronoi cells. Bottom panel: `Delaunay method'. The function $f$ is
approximated by linear interpolation from $f(x_i)$ to $f(x_{i+1})$ in
the interval $(x_i,x_{i+1})$, and similarly between $f(x_i)$ and
$f(x_{i-1})$ in the interval $(x_{i-1},x_i)$. The intervals
$(x_i,x_{i+1})$ and $(x_{i-1},x_i)$ can be considered as the
one-dimensional equivalents of Delaunay cells.}
\endinsert

Indeed the Voronoi method can be viewed as an elementary zeroth-order
interpolation scheme. However, it is possible to define a velocity field
based on linear interpolation between the velocities
of the sample points. This is the multidimensional equivalent
of the one-dimensional situation where the approximation of a function
by constant function values in bins centered on the sample points
(central panel figure 4) is replaced by linearly
interpolated function values in between those sample points (lower
panel figure 4). For $d=1$ linear interpolation simply consists of the
approximation of a function $f(r)$ by the value
$f(r)=\alpha_{i}\,f(r_i) \,+\, \alpha_{i+1}\,f(r_{i+1})$ in the
interval $r_i \leq r \leq r_{i+1}$, where $0 \leq \alpha \leq 1$ and
$\alpha_i + \alpha_{i+1} =1$. The natural extension to a space of
arbitrary dimension $d$ of the concept of the one-dimensional
interpolation interval $r_i \leq r \leq r_{i+1}$ is a `hyper-triangle'
defined by $d+1$ vectors ${\bf r}_k$, the vertices of that object. For
$d=2$ this is a triangle, for $d=3$ a tetrahedron. Any vector ${\bf r}$ within
the `hyper-triangle' is a linear combination of the $d+1$ vectors ${\bf
r}_k$, ${\bf r}\,=\, \sum\nolimits_{k=1}^{d+1}\, \alpha_{k} {\bf r}_k\,,$
with $0 \leq \alpha_k \leq 1$ and
$\alpha_1+\ldots+\alpha_{d+1}=1$. The linearity of the approximation
of the function $f({\bf r})$ then implies that

$$\eqalign{f({\bf r})\,&=\,f({\bf r_1})\,+\,\alpha_2\,\nabla f \cdot
({\bf r_2}-{\bf r_1}) \,+\, \ldots \,+\,\cr
&\qquad\qquad\qquad\qquad\qquad+\, \alpha_{d+1}\,\nabla f \cdot
({\bf r}_{d+1}-{\bf r}_1)\cr
&=\,f({\bf r}_1)\,+\,\alpha_2\,(f({\bf r}_2)-f({\bf r}_1))\,+\,\ldots\,+\,\cr
&\qquad\qquad\qquad\qquad\qquad+\alpha_{d+1}\,(f({\bf r}_{d+1})-f({\bf r}_1))
\,\cr
&=\, \sum_{k=1}^{d+1}\, \alpha_{k} f({\bf r}_k)\,.\cr}\eqnew$$
\@The choice of appropriate, and if possible optimal, interpolation
`hyper-triangles' is a critical issue that to a considerable extent
determines the quality and accuracy of the multidimensional linear
interpolation. An obvious minimal requirement is that the linear
approximation of the function $f$ leads to a {\it unique} value at
{\it every} point in the subspace
defined by the set ${\cal P}$. In other words, we need a space-filling
covering by mutual disjunct tetrahedra (for now we will restrict
ourselves to $d=3$, although it is trivial to follow the same argument
for any other $d$). In the interior of each of these tetrahedra each of
the velocity gradients $\partial v_i/\partial x_j$ has one particular
constant value, whose value is determined by the value of the function
$f$ at the locations ${\bf r}_k$ of its four vertices. A crucial
requirement for an optimal accuracy of the linear approximation
(eqn. 11) is that the tetrahedra in the space-filling network are
as compact as possible, in the sense of having a size and
elongation that are as small as possible. A uniquely defined solution
for such an optimal `triangulation' does not exist or, rather, is not
really known. Here we argue that a {\it Delaunay tessellation} (Delone
1934) is at least a good approximation of such an optimal triangulation.

Formally, the Delaunay tessellation ${\cal D}({\cal P})$ of a point
set ${\cal P}$ is defined to be the tessellation consisting of all the
tetrahedra defined by four nuclei whose circumscribing sphere is
{\it empty} in the sense that no nucleus of the generating set of nuclei
should be inside the circumsphere. A two-dimensional illustration of a
Delaunay triangulation is given in figure 1 (left-hand frame).
A principal characteristic of the circumsphere of a Delaunay
tetrahedron is that the centre of the circumsphere ({\it circumcentre})
is a vertex of the Voronoi tessellation. This follows immediately from the
extrapolation of the definition. After all, according to the
definition of the Voronoi tessellation a vertex is defined by four nuclei
that are equidistant from the vertex, i.e. the vertex is the circumcentre
of the circumsphere of these four nuclei. If then there were a
fifth nucleus within the sphere, it would be nearer to the
circumcentre than the four nuclei on the surface of the sphere. This
would imply that the centre cannot be the common vertex of the Voronoi
polyhedra of these four nuclei. Ergo, the four nuclei have to define a
{\it Delaunay tetrahedron}. In other words, the Delaunay tessellation
is the network that is obtained by joining all pairs of nuclei in
${\cal P}$ whose Voronoi polyhedra share a Voronoi wall. Such a pair
of nuclei is called a {\it contiguous pair}. The close
relationship between the Delaunay and the Voronoi tessellation can be
quite well appreciated from the right-hand frame of Figure 1,
depicting the Voronoi tessellation of the same point set as the one in the
left-hand frame.

Besides the fact that Delaunay tetrahedra fulfil the requirement of
compactness, them being objects of minimal size and elongation, an
additional and equally important argument to the use of the Delaunay
tetrahedra as the choice for linear interpolation intervals is
provided by the duality between Delaunay and Voronoi tessellations. Linear
interpolation requires the definition of neighbour intervals. An
arguably natural definition of `neighbour points' in the
multidimensional situation is that the two points share a Voronoi
wall, i.e. they should be contiguous to each other. Delaunay
tetrahedra therefore seem to be a natural choice for linear
interpolation intervals defined by four nuclei.

\topinsert
\vbox{\noindent\eightpoint
{\bf Figure 5.}
A visual impression of
the Voronoi (left-hand panel) and the Delaunay method (right-hand panel) for
approximating a function $f({\bf r})=f(r_1,r_2)$ of two variables $r_1$ and
$r_2$. Left-hand panel: the Voronoi method. The function $f$ has been
measured at the location of the black dots. In the $(r_1,r_2)$ plane the
corresponding Voronoi cells (neighbours of each other) have been
indicated by solid
lines. The resulting function field is indicated by the lightly shaded
polygons, the Voronoi cells around the central nuclei $r_i$. Each of
these polygons is positioned at the corresponding height $f(r_i)$. For
comparison with the Delaunay method, the darkly shaded triangle
indicates the value of the function field inside one of the
corresponding Delaunay triangles. Right-hand panel: the Delaunay method. The
dots
indicate the positions $r_i$ of the points at which the function
$f({\bf r}_i)$ has been determined. The same five Voronoi cells as in
the top panel are indicated by the solid lines in the $(r_1,r_2)$ plane, while
all the
Delaunay triangles corresponding to these cells have been indicated by
the dashed lines. The resulting function field is one of differently
inclined triangles (shaded), connecting with each other at the
boundaries of the Delaunay triangles. Evidently, at each of the
locations ${\bf r}_i$ (black dots) the function $f$ has the value
$f({\bf r}_i)$.}
\endinsert

Moreover, by virtue of
their duality, the Voronoi and the Delaunay tessellation for a given
point set ${\cal P}$ are calculated by the same three-dimensional
geometrical Voronoi tessellation code that was mentioned in the previous
subsection (Van de Weygaert 1991b, 1994). In this case the output
is limited to a listing of all vertices of the Voronoi tessellations and
the location of the four generating points that define the corresponding
Delaunay tetrahedron.

To give a visual impression of the relationship between the Voronoi
and the Delaunay approximation method, figure 4 shows the resulting
values of a function $f({\bf r})=f(r_1,r_2)$ for $d=2$. The Voronoi
method (left-hand frame) yields a field that consists of regions of a
constant value, Voronoi cells in $(r_1,r_2)$ plane. The resulting
image is therefore one of a field of pillars of different height, with
a Voronoi cell at the base of each of the pillars. The Delaunay
method divides space into triangular regions, the Delaunay triangles
in the $(r_1,r_2)$ plane, in which not the field value but the
field gradients are constant. This yields a field of differently
oriented triangles, connecting at their edges to the neighbouring
triangles (right-hand frame figure 5).

Having defined the Delaunay tetrahedra as the interpolation intervals,
we proceed by determining the values of the (constant) values of each
of the nine velocity gradient tensor components $\partial v_i /
\partial x_j$ in each of these tetrahedra. These are determined from
the location of each of the four vertices, ${\bf r}_0$,
${\bf r}_1$, ${\bf r}_2$ and ${\bf r}_3$, and the value of the velocity
field at each of these locations, ${\bf v}_0$, ${\bf v}_1$, ${\bf v}_2$ and
${\bf v}_3$. Defining the quantities $\Delta x_n \equiv x_n-x_0$, $\Delta y_n
\equiv y_n -y_0$ and $\Delta z_n \equiv z_n -z_0$, for $n=1, 2, 3$, as
well as $\Delta v_{xn} \equiv v_{xn} - v_{x0}$, $\Delta v_{yn}
\equiv v_{yn} -v_{y0}$
and $\Delta v_{zn} \equiv v_{zn}-v_{z0}$, the following nine linear
relations are obtained, with $n=1,2,3$,
$$\eqalign{
\Delta v_{xn} &=
{\displaystyle \partial v_x \over \displaystyle \partial x} \Delta x_n +
{\displaystyle \partial v_x \over \displaystyle \partial y} \Delta y_n +
{\displaystyle \partial v_x \over \displaystyle \partial z} \Delta z_n\cr
\Delta v_{yn} &=
{\displaystyle \partial v_y \over \displaystyle \partial x} \Delta x_n +
{\displaystyle \partial v_y \over \displaystyle \partial y} \Delta y_n +
{\displaystyle \partial v_y \over \displaystyle \partial z} \Delta z_n\cr
\Delta v_{zn} &=
{\displaystyle \partial v_z \over \displaystyle \partial x} \Delta x_n +
{\displaystyle \partial v_z \over \displaystyle \partial y} \Delta y_n +
{\displaystyle \partial v_z \over \displaystyle \partial z} \Delta z_n\cr}
\,,\eqnew$$
\@From these equations we can easily infer that the components of
${\partial v_i / \partial x_j}$ can be calculated from
$$\eqalign{\left(\matrix{
{\displaystyle \partial v_x \over \displaystyle \partial x}\cr\cr
{\displaystyle \partial v_x \over \displaystyle \partial y}\cr\cr
{\displaystyle \partial v_x \over \displaystyle \partial z}\cr}\right)
\,&=\,{\bf A}^{-1}\,
\left(\matrix{\Delta v_{x1}\cr\cr \Delta v_{x2}\cr\cr \Delta v_{x3}}\right);\cr
\left(\matrix{
{\displaystyle \partial v_y \over \displaystyle \partial x}\cr\cr
{\displaystyle \partial v_y \over \displaystyle \partial y}\cr\cr
{\displaystyle \partial v_y \over \displaystyle \partial z}\cr}\right)
\,&=\,{\bf A}^{-1}\,
\left(\matrix{\Delta v_{y1}\cr\cr \Delta v_{y2}\cr\cr \Delta v_{y3}}\right);\cr
\left(\matrix{
{\displaystyle \partial v_z \over \displaystyle \partial x}\cr\cr
{\displaystyle \partial v_z \over \displaystyle \partial y}\cr\cr
{\displaystyle \partial v_z \over \displaystyle \partial z}\cr}\right)
\,&=\,{\bf A}^{-1}\,
\left(\matrix{\Delta v_{z1}\cr\cr \Delta v_{z2}\cr\cr \Delta v_{z3}}\right)
\,,\cr}\eqnew$$
\@where ${\bf A}^{-1}$ is the inverse of the matrix
$${\bf A}\,=\,\left(\matrix{
\Delta x_1&\Delta y_1&\Delta z_1\cr\cr
\Delta x_2&\Delta y_2&\Delta z_2\cr\cr
\Delta x_3&\Delta y_3&\Delta z_3\cr}\right)\,,\eqnew$$
\@Note that the four points of the interpolation tetrahedron are
both necessary and sufficient to fix the value of each of the 9
velocity gradients. While in the linear regime only 6 of these
quantities would be needed, by virtue of the absence of vorticity, all
9 quantities are necessary in the nonlinear regime. From the values of
$\partial v_i / \partial x_j$ we can then easily
evaluate the value of the velocity divergence $\theta$, the shear
$\sigma_{ij}$ and the vorticity ${\bf \omega}$ (see eqn. 10) in each Delaunay
tetrahedron.

Subsequently we have to determine the corresponding volume averaged
quantities. As described in the former section we accomplish this
by top-hat filtering with a filter $W_{TH}$ that has a characteristic
radius $R$. The problem has therefore been reduced to determining the
average value of $\theta$, $\sigma_{ij}$ or ${\bf \omega}$ in a sphere
of radius $R$, centered on some location ${\bf x}_0$. For example, for the
volume averaged velocity divergence, $\tilde \theta ({\bf x})$, we have
$$\tilde\theta({\bf x}_0) \equiv {\displaystyle
\,\int d{\bf x}\,\theta({\bf x}) W_V({\bf x},{\bf x}_0) \over
\displaystyle \int d{\bf x}\,W_V({\bf x},{\bf x}_0)}\,.\eqnew$$
\@As the value of $\theta$ is constant within each
cell of the space-filling Delaunay tessellation, the problem reduces
to simply determining the intersection of the Delaunay tetrahedra with the
filter sphere (appendix B), and using the intersection volume as the
weighting value of $\theta$ in integral of equation (15). In other words,
$\tilde \theta$ follows from,
$$\tilde \theta ({\bf x}_0)={\displaystyle 3 \over \displaystyle
4 \pi R^3}\,\sum_k\,\theta_k \,V_{k,R}\,,\eqnew$$
\@where the sum is over all Delaunay tetrahedra $k$ that intersect
with the filter sphere, and $V_{k,R}$ is the volume of the
corresponding intersections, while $\theta_k$ is the value of $\theta$ in
$k$. It is good to note that the geometrical issue of determining the
intersection between a tetrahedron and a sphere is far from trivial. A
description of our algorithm is provided in Appendix B, but we should
note that more efficient evaluations are probably possible.
Evidently, the discussion is exactly analogous for the
volume-averaged values of the shear tensor components, $\tilde \sigma_{ij}$
of the vorticity, $\tilde \omega_{ij}$.

As with the Voronoi method, the final end result of the operations described
above consists of a field of top-hat averaged quantities at regular grid
intervals.

\medskip
\@{\bf {\bigone 4. Computational Considerations}}
\smallskip
In the former section we developed the theoretical groundwork for the use
of the Voronoi and the Delaunay tessellations as optimal tools for
determining the velocity divergence field, as well as the
vorticity and shear fields, from the value of the velocity field at
a discrete number of locations. The practical implementation of both
methods consists of a two-phase approach. (1) In the first step, the
algorithm calculates the Voronoi or Delaunay tessellation for the set
of points at which the velocity is known/measured. From the Voronoi
or Delaunay networks we can then define the value of the velocity
divergence, vorticity and/or shear throughout the whole of the sample volume.
(2) The subsequent second phase is the volume filtering of the
resulting tessellation defined field. In essence, volume filtering
consists of the determination of the average value of the velocity
gradient field, produced by the tessellation algorithm, within the
filter volume. Usually, the filtering averages are determined for a
set of regular grid positions.

\medskip
\@{\bf {4.1 The tessellation algorithm}}
\smallskip
On the basis of the duality between Voronoi and Delaunay tessellations
we use the same tessellation algorithm for both the Voronoi and the
Delaunay method. It is the geometric Voronoi tessellation code that was
developed by Van de Weygaert (1991b, 1994). For a very extensive and
detailed description of the code we refer to both these
references. Here we limit ourselves to a short overview of the
fundamental ideas.

The central operation of the geometric Voronoi algorithm is the
calculation of all the Delaunay tetrahedra of the point distribution.
Evidently, this simultaneously yields all Voronoi vertices, the centres of the
corresponding circumscribing spheres. Finding the appropriate
connections between the vertices, such that they define the edges, the
polygonal Voronoi walls and finally the complete polyhedral Voronoi
cells, is the major and most cumbersome part of the work. Our
algorithm is an elaboration on the work of Tanemura et al. (1983), who
gave a sketch of the basic algorithm for finding a Delaunay
tetrahedron and a Voronoi polyhedron, in combination with
four underlying geometric theorems. Through efficient administrative,
data storage and searching procedures our algorithm manages to
restrict itself to calculating every Delaunay tetrahedron and Voronoi
wall only once during the construction of the complete tessellation.
In a brute force construction procedure wherein each Voronoi cell is
completely calculated without use of prior information this would
be repeated three and one time respectively. This has
the additional advantage that it allows to discard a point $i$ from
any further consideration once its Voronoi cell $\Pi_i$ has been calculated.

For the sake of clarity it is good to recapitulate some basic properties of
Voronoi tessellations (also see Fig. 2). A nucleus $i$ defines
a complete polyhedral Voronoi cell. The boundary of the cell consists
of a set of polygonal walls, each of which is shared with a nucleus $j$ that
is contiguous to $i$. A Voronoi wall is the part of space that is as
close to $i$ as to $j$, and closer to these two nuclei than to any of
the other nuclei. On average there are $\approx 15.54$ walls per Voronoi
cell if the generating point distribution is Poissonian, an average
number that can vary from realization to realization (unlike the 2-D
case, where the average number of edges is always 6). The boundary of
a polygonal wall is formed by a number of edges, with a Voronoi vertex
at the two tips of each edge. In other words, the structure of a wall
is completely determined by listing the locations of the Voronoi
vertices along its edge in the right geometrical order. Note that each
edge is shared by three contiguous nuclei $i$, $j$ and $k$, to which every
point on the edge is equidistant while being closer to any of these three
than to any of the other nuclei. Likewise, a Voronoi vertex is
equidistant to its closest four nuclei in the generating point
process. For a Poisson point process there are $\approx 27.07$
vertices per Voronoi cell, and some $5.228$ per Voronoi wall.

Our code assumes that the generating point processes, and consequently
also the tessellations themselves, have periodic boundary conditions.
This assumption, however, can be relaxed with some additional
effort. Usually the points are distributed within a box of equally
sized edges, although any rectangular parallelepiped is feasible.
The first step of the algorithm is the storage of the set of generating
nuclei in a multidimensional binary tree. This data structure, on
average, stores points that are close together in space at nearby
positions of a binary tree structure (see e.g. Bentley 1986). Nearby
points in space can then
be tracked efficiently by means of a recursive searching procedure.
Building this tree is an $N \log N$ procedure, finding the nearest $M$
points is a task that requires a limited effort proportional to $M \log
N$ (see appendix F in Van de Weygaert 1994).

The algorithm then proceeds with the sequential computation of each
Voronoi cell $\Pi_i$. In turn, the construction of a cell $\Pi_i$
consists of the determination of the structure of each of its Voronoi
walls. A slight complication is of course that beforehand it is
unknown with which and with how many nuclei $i$ shares a wall. The identity
of these contiguous nuclei can be revealed in different
ways. Firstly, a subsample $S_i$ of $M \approx 50$ nearest nuclei is
selected. As contiguous points are likely to be close in physical
space, the subset $S_i$ is considered to be the primary sample of
contiguity candidates. Initially any search for a new contiguous
nucleus is restricted to $S_i$. If $i$ was not yet part of a Delaunay
tetrahedron, the first contiguous nucleus is the nucleus $i_1$ that is
closest to $i$. Evidently, $i_1$ belongs to $S_i$. A subsequent second
contiguous nucleus $i_2$ is found by looking for the nucleus that
together with $i$ and $i_1$ yields the triangle of minimal
circumradius. Although almost always $i_2 \in S_i$, there are
occasional exceptions, and an appropriate correction procedure for
this has to be included. A subsequent third contiguous nucleus $i_3$ is
the one that together with $i$, $i_1$ and $i_2$ defines a Delaunay
tetrahedron, the first one of which $i$ is one of the defining
nuclei. As in the case of $i_2$, the search for $i_3$ is initially
restricted to $S_i$, but occasionally corrective action is needed and
the search extended to a limited set of nuclei outside $S_i$. This
corrective procedure is efficiently performed via the multidimensional
binary tree. On the other hand, $i$ can have been one of the four defining
nuclei of one or more previously determined Delaunay tetrahedra. If
so, we know that the other three nuclei of any of these tetrahedra
are contiguous to $i$. In both situations an initial set $C_i$ of
contiguous nuclei is produced. Subsequently, we work our way through
the list $C_i$, calculating one by one the corresponding walls. For each
nucleus $i_{\alpha}$ in $C_i$ at least one Delaunay tetrahedron is
known. These tetrahedra are the starting point of any further search.
They are ordered in the appropriate geometrical order, and the gaps are filled
in. Within the wall around $\{i,i_{\alpha}\}$ a tetrahedron
${\cal D}_{\beta,\gamma} \equiv \{i,i_{\alpha},i_{\beta},i_{\gamma}\}$
should connect to a tetrahedron ${\cal D}_{\beta,\delta}\equiv
\{i,i_{\alpha},i_{\beta},i_{\delta}\}$. If such a
tetrahedron ${\cal D}_{\beta,\delta}$ already exists its centre will
be the Voronoi vertex connecting to the one of ${\cal
D}_{\beta,\gamma}$. If not, we have to search for
the nucleus $i_{\delta}$. First in the subset $S_i$, if necessary
followed by a corrective procedure. In this way new Delaunay tetrahedra are
not computed at random, but always starting from previous knowledge of three
of the four defining nuclei. It will automatically yield a new
contiguous nucleus, and $i_{\delta}$ is added to the list $C_i$.
Continuing along the same direction, the polygonal wall gets
progressively mapped by the connecting Voronoi vertices, a process
which is completed once a tetrahedron connects to
${\cal D}_{\beta,\gamma}$ on the side $\{i,i_{\alpha},i_{\gamma}\}$.
Meanwhile newly determined Delaunay tetrahedra have produced new
contiguous nuclei for the set $C_i$. After completion of the wall
around $\{i,i_{\alpha}\}$ the process proceeds with the next
contiguous nucleus in the list $C_i$. The construction of the
Voronoi polyhedron $\Pi_i$ has been completed once the walls around
all the nuclei in the constantly updated list $C_i$ have been determined.

In the construction procedure that we sketched in the previous
paragraph each Delaunay tetrahedron, and its corresponding Voronoi
vertex, is computed only once. Once four nuclei $i$, $i_{\alpha}$,
$i_{\beta}$ and $i_{\gamma}$ are found to constitute a Delaunay
tetrahedron, this information is passed on to all four of these
nuclei. Likewise with the Voronoi wall shared by the nuclei $i$ and $j$.
During construction of the Voronoi cell $\Pi_i$ the elaborate wall
construction procedure sketched above is skipped when the wall with
$j$ has been computed earlier during the construction of $\Pi_j$.
Instead, construction proceeds with the next nucleus in the
contiguity list $C_i$ for which not yet such a wall has been
determined, naturally only after having notified the cell $\Pi_i$ of
the existence of the wall it shares with $j$.
Besides selection of the subset $S_i$ of $M$ closest neighbours, one of the
first steps in the computation of a Voronoi cell $\Pi_i$ is therefore
processing of the information on the walls and vertices that were
obtained previously during the computation of other cells. This also
implies that the nucleus $i$ can be discarded from any further
considerations once the cell $\Pi_i$ has been determined, all of the
Delaunay tetrahedra in which it partakes already having been
calculated. This results in a progressive pruning of the
multidimensional tree during the tessellation construction procedure.

\medskip
\@{\bf {4.2 The filtering algorithms}}
\smallskip
The tessellation algorithms lead to constant values of the velocity
gradients within either the Voronoi walls, for the Voronoi method, or
Delaunay tetrahedra, for the Delaunay method. Subsequently, this field
is filtered (see Section 3.1 and 3.2). The operation of filtering is
basically equivalent to determining a weighted average of the field
within the filter volume. In the application of the Voronoi and the
Delaunay method we restrict ourselves to determining top-hat filtered
fields, which has the virtue that its filter value is a bounded sphere
of radius $R$.

Because in the Voronoi method the value of the velocity gradients
differs from zero only in the walls of the Voronoi tessellations,
the filtering consists of determining the surface area of the
intersections of the polygonal Voronoi walls with spheres of radius
R. The technicalities of this procedure are treated in Appendix A.

In the Delaunay method we end up with the more complicated situation
of constant non-zero values of the velocity gradients in the Delaunay
tetrahedra. The basic operation of top-hat filtering is therefore
determining the volume of the intersections of tetrahedra with a
sphere. This turns out to be a far from trivial geometrical problem.
In Appendix B we describe a provisional recipe to accomplish this
operation. However, it is quite likely that considerably more
efficient ways are feasible, which would lead to a substantial gain
in efficiency of the Delaunay method.

\medskip
\@{\bf {4.3 Computational effort}}
\smallskip
In practice, the amount of CPU time for the calculation of a
Voronoi cell in the Voronoi tessellation construction procedure is almost
independent of the number $N$ of generating nuclei, so that the
total tessellation construction time is almost linearly proportional
to $N$. The construction of the multidimensional binary tree is a
procedure that demands $\kappa N \log N$ time. In addition, a time $\propto
\lambda M \log N$ is required for the selection of each subset $S_i$
of $M$ closest neighbours requires, so that a total time
$\lambda (NM) \log N$ is spent on selecting all sets $S_i$.
It is the subsequent step for the actual
construction of each polyhedral Voronoi cell $\Pi_i$ that dominates the
CPU consumption. It is almost independent of $N$,
due to the fact that it takes place in subsets $S_i$ that contain
approximately the same number of $M \approx 50$ nuclei. Its total CPU
requirement is therefore something like $\mu N$, where $\mu \gg \lambda \simeq
\kappa$.

To illustrate the above, we quote some CPU tests on an IBM
7011/25T Power PC. For a range of 100 to 50,000 generating nuclei the
CPU time per cell is indeed constant, $\approx 0.010$ sec. For a
tessellation of 40,000 nuclei the tessellation computation time is
$\approx 407.28$ sec, while the corresponding tree building
time is $\approx 28.03$ sec. For comparison, in the case of a
tessellation of 1000 Voronoi cells this is $10.05$ sec and $0.25$ sec
respectively.

The construction of the Voronoi and Delaunay tessellation is
considerably less time
consuming then the ensuing Voronoi wall or Delaunay tetrahedron
intersection procedures. The major share of the effort lies
in the computation of the intersections between the spheres and
either the Voronoi walls or Delaunay tetrahedra. The CPU time
per intersection is basically constant. Here we quote some
numbers for a top-hat filter of $R=15.0h^{-1}\,\hbox{Mpc}$,
where the filtered quantities were determined at $20^3$ grid
positions in a box with an edge size of $200.0h^{-1}\,\hbox{Mpc}$,
containing a tessellation generated by $N=40,000$ nuclei. The number
of intersection evaluations is proportional to $R^3$, as well
as linearly proportional to $N$. With respect to the latter it
is worthwhile to realize that a Voronoi tessellation generated by $N$
nuclei has approximately $7.768 N$ Voronoi walls and $6.768N$ Delaunay
tetrahedra (i.e. $\approx 310,720$ walls and $\approx 270,720$
Delaunay tetrahedra for a Voronoi tessellation of $40,000$ cells).

In the case of the Voronoi method we need on average $0.126$ msec per
intersection. The surface area of approximately 5,759,500
intersections had to calculated, which resulted in a total CPU time
of $1454$ sec. As expected, an intersection between a sphere and
a Delaunay tetrahedron is more time demanding, taking $\approx 1.32$
msec, while the subsequent determination of the velocity gradient
tensor (eqn. 13) takes on average another $\approx 0.157$ msec. Our code
needed $12,882,264$ intersection determinations, so that some $17,034$
CPU seconds ($\approx 4.7$ CPU hours) were spent on this part of the procedure,
with an
additional $1235$ seconds for the velocity gradient tensor calculations.
Hereby we should add the remark that due to the way the code selects
the spheres and tetrahedra that are likely to intersect, some $39\%$
of the intersections actually turned out to be empty.

An additional important computational consideration concerns memory space.
Here the Delaunay method has a clear advantage over the
Voronoi method. For the Voronoi method we need to store all the
available information on the Voronoi tessellation. In this way we are
able to reconstruct the Voronoi walls, in particular the links between and
locations of the vertices that delineate those walls. As yet we still
use the output of the general Voronoi code, although we intend
to make a special-purpose version of the Voronoi code. The latter should
reduce the memory allocation considerably. For example, the structure
of the complete Voronoi tessellation of $40,000$ cells is contained in
7 files of in total 105 Mbyte. By contrast, the Voronoi code was
adapted for the Delaunay method so that only the information on the
Delaunay tetrahedra is stored, comprising 2 files of in total 14.3
Mbyte.

In practice the limitations of memory space are considerably more
restrictive than the required CPU time for the amount of data points
that can be handled by the Voronoi and the Delaunay tessellation
methods. For example, in the case of a data set of $128^3$ particles,
the Delaunay method would require a feasible $750$ Mbyte of memory
space. However, the more than 5.5 Gbyte needed by the Voronoi method
probably prohibit this method of becoming applicable to data sets
comprising more than $\approx 100,000$ points.

\medskip
\@{\bf {\bigone 5. Application: Velocity Statistics of an N-body
simulation}}
\smallskip
In order to compare the new `Voronoi tessellation method' (section 3.1), the
new `Delaunay tessellation method' (section 3.2) and the old
`two-step method', which we shortly described in section 2 (see e.g.
Juszkiewicz et al. 1995, Bernardeau 1994b), we have applied the new methods to
the result of an N-body simulation. This simulation, kindly provided
by H. Couchman, was obtained with an adaptive P$^3$M code (Couchman 1991)
with CDM initial condition
for $H_0=50$Mpc/km/s, and periodic boundary conditions in a cubic
box of 200$h^{-1}$Mpc size. The simulation consists of $128^3$
particles and has been evolved until the epoch when the r.m.s. linear density
fluctuations reaches unity in a spherical top-hat box of radius $8h^{-1}$Mpc.

\topinsert
\vbox{\noindent\eightpoint
{\bf Figure 6.}
Comparison of
the estimated values of the velocity divergence $\theta$ at the
present epoch, top-hat filtered with a filter radius
$R=15h^{-1}\,\hbox{Mpc}$, at $20^3$ grid
positions. The simulation consists of $128^3$ particles in a periodic
box of $200h^{-1} \hbox{Mpc}$ size, and concerns an $\Omega=1$
Universe with CDM initial conditions (see text). The left-hand panel is
a scatter plot of the value of $\theta$ at each of the grid
positions obtained with the two-step method against the value at the same
position for the Delaunay method. The right-hand panel is a similar
scatter plot, but then for the values obtained with the Voronoi method
and the Delaunay method.}
\endinsert

\topinsert
\vbox{\noindent\eightpoint
{\bf Figure 7.}
Scatter plots of the
value of the top-hat filtered velocity divergence $\theta$
determined by the Voronoi method against the value
determined by the Delaunay method, for $20^3$ grid positions. The plot
concerns the same $128^3$ CDM N-body simulation as in figure 6.
Each frame represents a different top-hat filter radius. Left-hand panel:
$R=2.5h^{-1}\,\hbox{Mpc}$. Central panel: $R=5.0h^{-1}\,\hbox{Mpc}$.
Right-hand panel: $R=10.0h^{-1}\,\hbox{Mpc}$}
\endinsert

As it was not feasible for us to apply the Voronoi method for
a particle sample of $128^3$ points (see discussion at the end
of section 4), we needed to reduce the number of particles used
in the analysis. To do so we take advantage of the fact that
the filtered velocity field at the radius corresponding to the
quasi linear regime is not sensitive to the details of the
small scale velocity field in the very dense regions. Therefore,
the statistical quantities we are interested in are not affected if we
sample the data set in such a way that it does not reduce
the number density of particles in the underdense regions while it does
so in the very dense areas. To achieve this goal we have constructed an
algorithm in which eight different grids are superposed, all
with the same grid size but shifted by half of the grid-size in each of the
possible directions. Subsequently, we perform a sequential operation on the
set of simulation particles, whereby we consider the location of each
particle, rejecting
those whose locations in {\it all} of the shifted grids would be in a grid-cell
that has already been occupied by a previously considered particle.
For a grid size of $10h^{-1}$ Mpc the final number of points
is about 40000. It corresponds to a mean distance between points
of about 6 $h^{-1}$Mpc. We constructed the Voronoi and Delaunay tessellations
of these particle samples, in order to derive the statistical properties of
the velocity field in the way that was described in details in the previous
section. In addition, we checked the robustness of the results by repeating
the same analysis for a grid-size of $7.5 h^{-1}$Mpc and double amount of
points.

\bigskip
\centerline{\hbox to 450pt{\bf Table 1. Statistical parameters
for the density and the velocity field at different radii \hfill}}
\smallskip
\centerline{\vbox{\tabskip=0pt \offinterlineskip
\def\tablerule{\noalign{\hrule}}
\hrule
\halign to450pt{\vrule#\tabskip=1em plus2em&\hfil\strut$#$\hfil&
\vrule$#$&\hfil\strut$#$\hfil&\vrule$#$&\hfil\strut$#$\hfil&\vrule$#$&
\hfil\strut$#$\hfil&\vrule$#$&\hfil\strut$#$\hfil&\vrule$#$\tabskip 0pt\cr
height8pt&\omit&&\omit&&\omit&&\omit&&\omit&\cr
&{\rm Radius}/(h^{-1}{\rm Mpc})
&&$7.5$&&$10.$&&$12.5$&&$15.$&\cr
height8pt&\omit&&\omit&&\omit&&\omit&&\omit&\cr
\tablerule
height8pt&\omit&&\omit&&\omit&&\omit&&\omit&\cr
&n&&-1.16&&-0.98&&-0.86&&-0.73&\cr
height3pt&\omit&&\omit&&\omit&&\omit&&\omit&\cr
&\gamma_2&&-0.51&&-0.57&&-0.60&&-0.64&\cr
height8pt&\omit&&\omit&&\omit&&\omit&&\omit&\cr
&\sigma_{\delta}&&0.99&&0.74&&0.57&&0.47&\cr
height3pt&\omit&&\omit&&\omit&&\omit&&\omit&\cr
&\sigma_{\theta}^{\Del.}&&0.63\pm0.02&&0.53\pm0.02&&0.45\pm0.015&
&0.38\pm0.015&\cr
height3pt&\omit&&\omit&&\omit&&\omit&&\omit&\cr
&\sigma_{\theta}^{\Vor.}&&0.67\pm0.02&&0.55\pm0.02&&0.46\pm0.015&
&0.39\pm0.015&\cr
height8pt&\omit&&\omit&&\omit&&\omit&&\omit&\cr
&-T_3&&1.87&&1.69&&1.57&&1.44&\cr
height3pt&\omit&&\omit&&\omit&&\omit&&\omit&\cr
&-T_3^{\Del.}&&1.74\pm0.06&&1.66\pm0.06&&1.58\pm0.10&&1.50\pm0.12&\cr
height3pt&\omit&&\omit&&\omit&&\omit&&\omit&\cr
&-T_3^{\Vor.}&&1.70\pm0.05&&1.64\pm0.07&&1.56\pm0.09&&1.48\pm0.12&\cr
height8pt&\omit&&\omit&&\omit&&\omit&&\omit&\cr
&T_4&&5.35&&4.04&&3.25&&2.47&\cr
height3pt&\omit&&\omit&&\omit&&\omit&&\omit&\cr
&T_4^{\Del.}&&4.9\pm0.6&&4.6\pm0.6&&4.5\pm1.5&&4.5\pm2&\cr
height3pt&\omit&&\omit&&\omit&&\omit&&\omit&\cr
&T_4^{\Vor.}&&4.6\pm0.4&&4.3\pm0.9&&4.2\pm1.5&&4.3\pm2&\cr
height8pt&\omit&&\omit&&\omit&&\omit&&\omit&\cr}
\hrule
}}
\bigskip

\medskip
\@{\bf{5.1 Results for the velocity divergence}}
\smallskip
In order to illustrate the substantial improvements that are obtained
by the Voronoi tessellation method and the Delaunay tessellation method,
in comparison with older methods like the two-step one, and to compare the
results to known theoretical predictions, we will in particularly focus
our analysis on the velocity divergence. The same simulation was tentatively
analyzed with the two-step method by Bernardeau (1994b, also see Juszkiewicz
et al. 1995). In the scatter plot of figure 6 we compare the estimates of
$\theta$ that we obtained with the Delaunay method on $20^3$ locations on a
regular grid, with those obtained by the two-step method (left-hand frame)
and the Voronoi method (right-hand frame), both on the same grid locations.
The scatter plat clearly shows the good agreement between the Voronoi and the
Delaunay tessellation method. Given the fact that both methods are quite
different, this provides considerable confidence in them. On the other
hand, the old two-step method yields very noisy estimates. Moreover, it tends
to underestimate the value of $\theta$ by a factor of about 1.2. This effect
is probably due to the fact that the effective smoothing radius is slightly
larger because of the combination of the two smoothing procedures.

In figure 7 we present similar scatter plots, for three different smoothing
radii, to show the correlation between the divergences measured by the
Voronoi and the Delaunay methods. The noise becomes very important for radii
smaller than $5h^{-1}$Mpc, that is when the radius becomes smaller than
the mean distance of the sampled particles. The results obtained for
a smoothing length of $2.5h^{-1}$Mpc are obviously not reliable
and show specific features associated with the methods that
have been used. For example, at these small radii a large fraction of the
smoothing spheres do not intersect any of walls of the Voronoi tessellation.
The Voronoi method therefore yields values of zero for $\theta$ in these
spheres. This situation is actually rather similar to the case of Poisson
noise, with the velocity divergence only having a non-zero value at some
discrete locations. For the Delaunay method the effect is less
dramatic. However, the measured velocity divergence gets affected
by the fact that information from larger scales, i.e. from the mean
separation scale, leaks into the local velocity. This effect can also be
traced in the other scatter plots.  In the latter we note that the measured
values of the velocity divergence tend to be smaller in the case of
the Delaunay method than
in the case of Voronoi method. Overall, however, at a scale of
10-15$h^{-1}$Mpc we expect all measured quantities to be free of systematic
errors, so that they can be analyzed with confidence.

In Table 1 we summarize the resulting values of the moments of
the velocity divergence obtained with the
Delaunay method  (with superscript $^{\Del.}$) and the Voronoi method
(with superscript $^{\Vor.}$), from an analysis in which the
velocity divergence has been measured at $50^3$ different
grid points. The error bars have been obtained by dividing the simulation in
four equal subsamples, and performing the same analysis for each of the
subsamples. The results on the variance clearly show
that the r.m.s. of the velocity divergence, $\sigma_{\theta}$,
is lower than the r.m.s. density fluctuations, $\sigma_{\delta}$.
This effect was also noticed by Juszkiewicz et al. (1995), and is an
indication
for the departure of the dynamics from the linear approximation, since
in the linear regime $\sigma_{\theta}$ and $\sigma_{\delta}$ are expected
to be equal.

\topinsert
\vbox{\noindent\eightpoint
{Figure 8.}
The probability distribution
function (PDF) of the top-hat filtered velocity divergence $\theta$
at the present epoch, for the same CDM $128^3$ particle N-body
simulation as in figure 6 and 7, determined from the values on a
$50^3$ grid (see text). Top panel: filter radius $R_{TH}=10h^{-1}\,
\hbox{Mpc}$.
Bottom panel: filter radius $R_{TH}=15h^{-1}\,\hbox{Mpc}$. The solid
lines represent theoretical predictions of the PDF for the measured
values of $\sigma_{\theta}$ (Bernardeau
1994b). The top panel one is given by eqn. (22), see Bernardeau
(1994b), while the one in the lower panel has been obtained
by numerical integration (see Bernardeau 1994b). Black squares: the
Voronoi method. Black triangles: the Delaunay method. Open squares:
two-step grid method.}
\endinsert

However, our main interest concerns higher order moments. In the past years
a lot of theoretical results on the statistical properties of the
velocity divergence have been obtained. Bernardeau (1994a) derived the
following results on the third and fourth order moment for
cosmological models with Gaussian initial conditions, $\Omega=1$ and
$\Lambda=0$. For the condition of a small $\sigma_{\theta}$, the following
expressions were found,
$${\mg\theta^3\md\over \mg\theta^2\md^2}=T_3+O(\sigma_{\theta}^2(R))\eqnew$$
\@with
$$T_3={26\over 7}+\gamma_1\,,\eqnew$$
\@and
$${\mg\theta^4\md-3\mg\theta^2\md^2\over \mg\theta^2\md^3}
=T_4+O(\sigma_{\theta}^2(R))\eqnew$$
\@with
$$T_4={12088\over 441}+{338\over21}\,\gamma_1+{7\over3}\gamma_1^2+{2\over3}
\gamma_2\,,\eqnew$$
\@where the parameters $\gamma_1$ and $\gamma_2$ are the successive
logarithmic derivatives of $\sigma_{\theta}^2(R)$ with respect to $R$,
$$\eqalign{
\gamma_1&={d\log\sigma_{\theta}^2(R)\over d\log R}\equiv-(n+3),\cr
\gamma_2&={d^2\log\sigma_{\theta}^2(R)\over d\log^2 R}.}\eqnew$$
\@For a CDM spectrum, Table 1 lists the values of the parameters $n$ and
$\gamma_2$ for different smoothing radii. The corresponding theoretical
values of $T_3$ and $T_4$ can be determined from equations (18) and
(20). The values for $T_3$ measured from the simulation by both the
Voronoi and the Delaunay method are found to be in remarkable agreement
with these theoretical predictions in the case of all four different radii.
While it was not possible to determine $T_4$ as accurately in the case of the
largest radius, it was found to agree reasonably well with the theoretical
predictions for radii $R \lessapprox 10h^{-1}$Mpc, i.e. at radii for which
this quantity could be measured sufficiently accurately. This solves the issue
raised by \L okas et al. (1995), who questioned the validity of the
perturbation theory for the divergence of the velocity field. From our
results we can conclude that the departure they observed in their work
was due to the systematic errors introduced by the smoothing schemes they used.

\medskip
\@{\bf {5.2 The Probability Distribution Function (PDF)
of the velocity divergence}}
\smallskip
Although it is interesting to study the individual moments, as they highlight
different features of the total distribution, a more complete
picture is obtained by looking at the global shape of the
PDF of the velocity divergence. In figure 8 we present, for two different
radii, the measured velocity divergence PDF. These measures are compared to
the theoretical predictions of Bernardeau (1994b) obtained from
the re-summation of the series of the cumulants (solid curves).
For $n=-1$ there is a simple analytical expression that can be
used for the PDF of $\theta$ (given here for $\Omega=1$),
$$\eqalign{&p(\theta){\rm d}\theta=\cr
&\ \ {([2\kappa-1]/\kappa^{1/2}+
[\lambda-1]/\lambda^{1/2})^{-3/2}
\over \kappa^{3/4} (2\pi)^{1/2} \sigma_{\theta}}
\exp\left[-{\theta^2\over 2\lambda\sigma_{\theta}^2}\right]
{\rm d}\theta,\cr}\eqnew$$
\@with
$$ \kappa=1+{\theta^2\over 9\lambda }\,,\qquad \hbox{and} \qquad
\lambda=1-{2\theta\over 3}\,.\eqnew$$
\@This expression was used in the plot for $R=10h^{-1}$Mpc. For
$R=15 h^{-1}$ Mpc, however, the solid curve was obtained by numerical
integration of the inverse Laplace transform (similar to Eq. 18 of
Bernardeau 1994b).

The PDFs measured by both the Voronoi and the Delaunay method are clearly
in remarkably good agreement with the theoretical predictions, down to
probabilities of about $10^{-4}$. On the other hand, previous methods, like
the two-step method (open symbols in Fig. 8) produce PDFs that deviate
substantially from the theoretical curves, producing spurious tails at
both the low and high value end of the PDF. In particular noteworthy
is the fact that the new tessellation methods manage to reproduce the
expected sharp cutoff at the positive values of $\theta$, which corresponds
to voids.

\bigskip
\centerline{\hbox to 475pt{\bf Table 2. The relative magnitude of
the divergence, vorticity and shear. \hfill}}
\smallskip
\centerline{\vbox{\tabskip=0pt \offinterlineskip
\def\tablerule{\noalign{\hrule}}
\hrule
\halign to475pt{\vrule#\tabskip=1em plus2em&\hfil\strut$#$\hfil&
\vrule$#$&\hfil\strut$#$\hfil&\vrule$#$&\hfil\strut$#$\hfil&\vrule$#$&
\hfil\strut$#$\hfil&\vrule$#$&\hfil\strut$#$\hfil&\vrule$#$\tabskip 0pt\cr
height8pt&\omit&&\omit&&\omit&&\omit&&\omit&\cr
&{\rm Radius}/(h^{-1}{\rm Mpc})
&&$7.5$&&$10.$&&$12.5$&&$15.$&\cr
height8pt&\omit&&\omit&&\omit&&\omit&&\omit&\cr
\tablerule
height8pt&\omit&&\omit&&\omit&&\omit&&\omit&\cr
&\sigma_{\theta}^{\Del.}&&0.63\pm0.02&&0.53\pm0.02&&0.45\pm0.015&
&0.38\pm0.015&\cr
height3pt&\omit&&\omit&&\omit&&\omit&&\omit&\cr
&\sigma_{\theta}^{\Vor.}&&0.67\pm0.02&&0.55\pm0.02&&0.46\pm0.015&
&0.39\pm0.015&\cr
height8pt&\omit&&\omit&&\omit&&\omit&&\omit&\cr
&\overline{\omega}^{\Del.}&
&0.183\pm0.003&&0.118\pm0.001&&0.081\pm0.001&&0.059\pm0.001&\cr
height3pt&\omit&&\omit&&\omit&&\omit&&\omit&\cr
&\overline{\omega}^{\Vor.}&
&0.243\pm0.004&&0.147\pm0.002&&0.098\pm0.0015&&0.070\pm0.001&\cr
height8pt&\omit&&\omit&&\omit&&\omit&&\omit&\cr
&\overline{\sigma}^{\Del.}&
&0.50\pm0.01&&0.41\pm0.01&&0.347\pm0.009&&0.297\pm0.008&\cr
height3pt&\omit&&\omit&&\omit&&\omit&&\omit&\cr
&\overline{\sigma}^{\Vor.}&
&0.55\pm0.01&&0.45\pm0.01&&0.366\pm0.009&&0.310\pm0.008&\cr
height8pt&\omit&&\omit&&\omit&&\omit&&\omit&\cr}
\hrule
}}
\bigskip

The accuracy of the perturbation theory calculations is therefore confirmed
for the shape of the PDF of the velocity divergence. This is clearly of
great importance, as the shape of the PDF can be potentially used to measure
$\Omega$ (Bernardeau et al. 1995).

\topinsert
\vbox{\noindent\eightpoint
{\bf Figure 9.}
Comparison of the values
of the top-hat averaged vorticity $\omega$ (eqn. 24) (left-hand frame)
and shear $\sigma$ (eqn. 25) (right-hand frame) for the same
CDM $128^3$ particle N-body simulation as in figures 6-8, for a top-hat
filter radius of
$R_{TH}=15h^{-1}\,\hbox{Mpc}$ (both $\omega$ and $\sigma$ in units of
$H$). Both frames represent scatter plots
of the values of these quantities at $20^3$ grid positions as
determined by the Voronoi method against those determined at the
same position by the Delaunay method.}
\endinsert

\medskip
\@{\bf {5.3 Vorticity and Shear}}
\smallskip
When introducing the Voronoi and Delaunay method in section 3, we made the
observation that both methods can actually be used to study the statistical
properties of any quantity related to the velocity deformation
tensor, such as the vorticity and the shear.
In figure 9 we compare by means of scatter plots the results that have
been found with our two methods for the norms of these
quantities, $\omega$ (left-hand frame) and $\sigma$ (right-hand frame),
$$\omega^2=\sum_k \omega_k^2,\eqnew$$
$$\sigma^2=\sum_{i,j} \sigma_{ij} \sigma_{ij}.\eqnew$$
\@These plots should be compared with the similar plot of the velocity
divergence in the right-hand frame of figure 6. Because the mean vorticity is
pretty small its statistics is quite sensitive to noise. Even for a radius as
large as $15h^{-1}$Mpc the resulting measured value of
the mean vorticity can be significantly affected by
systematics errors. Since it does not vanish in the linear regime such a
sensitivity does not exist for the velocity shear. A summary of
the expectation values of $\omega$ and $\sigma$ is given in table 2,
which also contains the values obtained for the r.m.s. of $\theta$.

\topinsert
\vbox{\noindent\eightpoint
{\bf Figure 10.}
Log-log plots of the
probability distribution functions (PDFs) of the vorticity $\omega$
(in units of $H$, left-hand frame) and shear $\sigma$ (in units of $H$,
right-hand frame) of the same
$128^3$ particle CDM N-body simulation as in Fig. 6-9, determined from
the values on a $50^3$ grid. Both frames concern the values of
these quantities smoothed with the same filter as in figure 9,
a top-hat filter with radius of $R_{TH}=15h^{-1}\,\hbox{Mpc}$.
Black squares: Voronoi method. Black triangles: Delaunay method.}
\endinsert

\topinsert
\vbox{\noindent\eightpoint
{\bf Figure 11.}
Scatter plots of the value
of the local density contrast $\delta$ against the velocity divergence
$\theta$ (left-hand frame), vorticity $\omega$ (central frame) and
shear $\sigma$ (right-hand frame), at $20^3$ grid positions, for the
same $128^3$ particle CDM N-body simulations as in Figs. 6-10. All
quantities are top-hat filtered with a top-hat radius $R_{TH}=15h^{-1}\,
\hbox{Mpc}$. The solid line in the left-hand panel indicates the
prediction of the $\delta-\theta$ relation by Bernardeau (1992).}
\endinsert

We find that the amount of shear is slightly smaller than the
amount of divergence. In fact, the value for $\sigma$ found with both
methods is almost exactly consistent with the magnitude of the shear
expected in the linear regime,
$${1 \over H^2}\,\langle\sigma^2\rangle\,=
\, {2 \over 3} \,\langle \theta^2 \rangle\,.\eqnew$$
\@As far as the amount of vorticity is concerned, we
see that it shows the expected rapid decrease with scale.
For $R\largapprox 10 h^{-1}$ Mpc it seems to be fair to assume that the
vorticity is small compared to the divergence. To give a crude idea of
them, figure 10 shows the PDF of $\omega$ and $\sigma$, for $R=15h^{-1}$Mpc,
obtained with the two new methods.

\medskip
\@{\bf {5.4 The local density-velocity relationship}}
\smallskip
In addition to the analysis of the statistical properties of the
velocity field, it is also possible to use the Voronoi and the
Delaunay method to study the joint distributions of the density and the
velocity field. Figure 11 displays scatter plots of the local density
contrast $\delta$ against respectively divergence (left-hand frame),
vorticity (central frame) and shear (right-hand frame). The strong
correlation between the density and the divergence is as expected, although
there is a quite a large amount of scatter. For comparison, the solid line
shows the prediction by Bernardeau (1992). From the scatter plot
against the vorticity we can infer that the mean vorticity increases
slightly with the density. A similar conclusion can be made for the shear.
It confirms the idea that voids tend to be regular spherically expanding
regions, whereas dense matter concentrations tend to have non-radial
motion.

\medskip
\@{\bf {\bigone 6. Summary and Discussion}}
\smallskip
The velocity field in the local Universe is an important and essential
source of information on structure formation. In particular
interesting for an understanding of the evolution and dynamics of the
structures in the Universe are the various components of the gradient
of the velocity field, the velocity divergence, shear and vorticity.
One approach is to study the statistical properties, both moments as
well as the full probability distribution function (PDF), of the divergence,
shear and vorticity of the local smoothed velocity field. Considerable
effort has been directed to obtaining analytical results for the
statistics of the velocity divergence in the linear and quasi-linear
regime in the case of structure formation scenarios based on Gaussian
initial density and velocity fields.

To study more advanced stages of structure evolution we often have
to resort to N-body simulations, yielding discretely sampled
velocity fields. The discrete nature of the velocity sampling
complicates the determination of the statistical properties
of the velocity field. In this paper we have introduced and
developed two numerical methods, the {\it Voronoi tessellation method}
and the {\it Delaunay tessellation method}, that yield reliable and
accurate estimators of volume-averaged quantities in the
case of discretely sampled velocity fields. The fact that they concern
volume-averaged quantities is of crucial importance. Almost all
analytical results
concern volume-average quantities while in essence all available
numerical estimators only yield mass-averaged quantities. The latter
considerably obscured the comparison between statistical results
from analytical models and N-body studies, and even lead to false conclusions
regarding e.g. the validity of perturbation theory.

The availability of estimators of volume-averaged velocity statistics
is important for several reasons. Firstly, it allows us to check
independently whether the perturbation calculations
that yield the quasi-linear results are indeed valid. Secondly,
if so, we can apply the new estimators with confidence to
highly nonlinear circumstances. And finally it may be feasible to
apply them, in adapted form, to the available catalogues of measured
galaxy peculiar velocities.

The {\it Voronoi tessellation method} and the {\it Delaunay
tessellation method}.
Both methods are based on two important objects in stochastic
geometry, the Voronoi and the Delaunay tessellation of a point set.
A Voronoi tessellation of a set of nuclei is a space-filling
network of polyhedral cells, each of which delimit the part of space
that is closer to its nucleus than to any of
the other nuclei. The Delaunay tessellation is also a space-filling
network of mutual disjunct objects, tetrahedra in 3-D. The four
vertices of each Delaunay tetrahedron are nuclei from the point set,
such that the corresponding circumscribing sphere does not
have any of the other nuclei inside. The Voronoi and the Delaunay
tessellation are closely related, and are dual in the sense that one
can be obtained from the other.

In a first evaluation of the two new methods we calculated, on a regular
grid, the volume-averaged velocity divergence, shear and vorticity of an
$128^3$ particle N-body simulation of structure formation in an
$\Omega=1$ CDM Universe. Computer memory space limitations in the case
of the Voronoi method forced us to sample some $40,000$ particles from
the total sample of $128^3$ particles. This sampling was performed
such that the number density in underdense regions was not reduced. A
comparison study between the Voronoi method, the Delaunay
method, the conventional `two-step' method and analytical theoretical
predictions yields encouraging results for our new methods. The
comparison study consists of comparison of scatter plots, third and
fourth order moments as well as the global PDFs. The Voronoi and the
Delaunay method show remarkable good agreement with each other, as
well as with theoretical predictions. On the other hand, considerable
differences with the conventional `two-step' method were found.

We may therefore conclude that the Voronoi and the Delaunay methods
represent optimal estimators for determining the probability
distribution function of volume-averaged velocity field quantities.
As yet it is difficult to judge which of the two methods is the
preferable one. The results produced by both methods agree very well for a
considerable range of situations. However, the Delaunay method is
clearly the better one for small filter radii, as it provides a
reasonable estimate
of the velocity gradients throughout the whole of space. The Voronoi
method, on the other hand, only does so in the Voronoi walls.
Consequently, irrelevant and noisy filter averages are produced by the
latter if the filter scale is smaller than the average wall distance
because the small filter spheres frequently end up being empty.
Another advantage of the Delaunay method is its approximately 8 times
lower memory space requirement. However, this may be a mere practical issue,
a more efficient Voronoi method implementation is certainly
feasible. A clear disadvantage of the Delaunay method is the
fact that it is almost 12 times more CPU time consuming than the
Voronoi method. This is for a large part due to the inefficient
calculation of the intersection between a tetrahedron and a sphere.
We expect that better and faster prescriptions are possible, which may
possibly lead to a five to tenfold acceleration of this algorithm.

In forthcoming work we will apply the newly developed tool to a plethora
of structure formation scenarios, based on both Gaussian as well as
non-Gaussian initial conditions. The reliability of the
results obtained with both the Voronoi and the Delaunay method allows us
to study in how far the velocity field PDFs are sensitive
discriminators that highlight physical differences between the scenarios. In
particular, we are interested in the possibility of extracting the
value of $\Omega$ from these velocity statistics. The availability
of the reliable numerical estimators that we developed here is a
crucial step in making this a practical possibility. We therefore also
wish to develop our methods for the even less ideal circumstances of
observational data. To see whether this is feasible, one of the first
steps is to see in how far our methods are sensitive to substantial
amounts of noise in the data. These issues will be addressed in a
forthcoming study. Also, we intend to make the software that we
developed for both the Voronoi and the Delaunay method publicly
available once it has been made user-friendly.

In addition, we feel that variations of the two numerical tools that we have
introduced here can be applied to a choice of other applications
in astrophysical situations. In many situations the value of a
particular physical quantity is only known at a limited number of discrete
points in space. The optimal adaptive nature
of both the Voronoi and the Delaunay tessellations to the point
distribution makes methods based on them promising estimators of
the general run of quantities over the whole of the sample space.

\medskip
\@{\bf {\bigone Acknowledgments}}
We are very grateful to Dick Bond, Christophe Pichon and Simon White
for useful and encouraging discussions and suggestions, and to Hugh Couchman
for providing the results of the CDM N-body simulation. In addition,
we would like to thank Bhuvnesh Jain for useful suggestions, and
Jens Villumsen for encouraging remarks on Voronoi tessellations. In
particular we are indebted to the referee, Edmund Bertschinger, for
useful suggestions and comments. FB is grateful for the hospitality of
the MPA, and RvdW for the hospitality of CE de
Saclay, where most of the work presented in this paper was carried out.

\bigskip
\@{\bf {\bigone References}}
\medskip
\hang Bentley, J.L., 1986, Programming Pearls, Addison-Wesley
\hang Bernardeau, F., 1992, ApJ, 390, L61
\hang Bernardeau, F., 1994a, ApJ, 433, 1
\hang Bernardeau, F., 1994b, A\&A, 291, 697
\hang Bernardeau, F., Juszkiewicz, R., Dekel, A., Bouchet, F.R., 1995, MNRAS,
in press
\hang Bertschinger, E., Dekel, A., 1989, ApJ, 336, L5
\hang Bertschinger, E., Dekel, A., Faber, S.M., Dressler, A., Burstein, D.,
1990, ApJ, 364, 370
\hang Burstein, D., Davies, R.L., Dressler, A., Faber, S.M., Stone, R.P.S.,
Lynden-Bell, D., Terlevich, R.J., Wegner, G.A., 1987, ApJS, 64, 601
\hang Coles, P., 1990, Nature, 346, 446
\hang Couchman, H.M.P., 1991, ApJ, 368, L23
\hang Dekel, A., 1994, ARAA, 32, 371
\hang Dekel, A., Rees, M., 1994, ApJ, 422, L1
\hang Delone, B.V., 1934, Bull. Acad. Sci. (VII) Classe Sci. Mat., 793
\hang Dirichlet, G.L., 1850, J. reine angew. Math., 40, 209
\hang Finney, J.L., 1979, J. Comput. Phys., 32, 137
\hang Gilbert, E.N., 1962, Ann. Math. Stat., 33, 958
\hang Goldwirth, D.S., Da Costa, L.N., Van de Weygaert, R., 1995, MNRAS,
in press
\hang Icke, V., van de Weygaert, R., 1987, A\&A, 184, 16
\hang Ikeuchi, S., Turner, E.L., 1991, MNRAS, 250, 519
\hang Juszkiewicz, R., Weinberg, D.H., Amsterdamski, P., Chodorowski, M.,
Bouchet, 1995, ApJ, 442, 39
\hang Kiang, T., 1966, ZfA, 64, 433
\hang \L okas, E.L., Juszkiewicz, R., Weinberg, D.H., Bouchet, F.R., 1995,
MNRAS, in press
\hang Matsuda, T., Shima, E., 1984, Prog. Theor. Phys., 71, 855
\hang Meyering, J.L., 1953, Philips Res. Rept., 8, 270
\hang Miles, R.E., 1970, Math. Biosc., 6, 85
\hang M{\o}ller, J., 1989, Adv. Appl. Prob., 21, 37
\hang Nusser, A., Dekel, A., 1993, ApJ, 405, 437
\hang Peebles, P.J.E., 1980, The Large-Scale Structure of the Universe,
Princeton Univ. Press
\hang Rubin, V.C., Thonnard, N., Ford, W.K., Roberts, M.S., 1976, AJ, 81, 719
\hang Smoot, G.F., Lubin, P.M., 1979, ApJ, 234, L83
\hang Stoyan, D., Kendall, W.S., Mecke, J., 1987, Stochastic Geometry and
Its Applications, Akademie-Verlag, Berlin
\hang Strauss, M.A., Davis, M., Yahil, A., Huchra, J.P., 1990, ApJ, 361, 49
\hang SubbaRao, M.U., Szalay, A.S., 1992, ApJ, 391, 483
\hang Tanemura, M., Ogawa, T., Ogita, N., 1983, J. Comp. Phys., 51, 191
\hang Van de Weygaert, R., 1991a, MNRAS, 249, 159
\hang Van de Weygaert, R., 1991b, Ph.D. thesis, Leiden University
\hang Van de Weygaert, R., 1994, A\&A, 283, 361
\hang Van de Weygaert, R., Icke, V., 1989, A\&A, 213, 1
\hang Villumsen, J.V., 1995, MNRAS, submitted
\hang Voronoi, G., 1908, J. reine angew. Math., 134, 198
\hang Williams, B.G., 1992, Ph.D. thesis, Edinburgh University
\hang Williams, B.G., Peacock, J.A., Heavens, A.F., 1991, MNRAS, 252, 43
\hang Yahil, A., Strauss, M.A., Davis, M., Huchra, J.P., 1991, ApJ, 372, 380
\hang Yoshioka, S., Ikeuchi, S., 1989, ApJ, 341, 16

\topinsert
\vbox{\noindent\eightpoint
{\bf Figure 12.}
The intersection of a sphere and a
polygon. The dashed disc is the intersection of the sphere with the
plane supporting the polygon. The points I, J are the intersecting
points of the polygon with the small circle.}
\endinsert

\bigskip
\@{\bf {Appendix A: The intersection of a sphere and a polygon}}
\smallskip
For the implementation of the Voronoi method we need to calculate the
surface of the intersection of a polygons and a sphere.

Figure 12 sketches the geometrical situation we encounter.
The problem naturally reduces to a planar problem
in the plane of the polygon (bottom panel) where one has
to calculate the surface of the intersection of the polygon with the
disc. This figure corresponds to the generic
case where the circle intersects the polygons in two
different points I, J. The surface is then given by the sum of the
surface of the polygon (I,A,B,C,D,J) and the surface of the
segment defined by the points I,J,
given by $\pi (\theta-\sin(\theta)) R_d^2/2$,
where $R_d$ is the radius of the disc.
There are in fact other possible cases: the disc may be entirely
contained in the polygon (the surface is then
the one of the disc), may contain the whole polygon (the surface is the
surface of the polygon) or
may have more than two intersection points with the polygon.
In the latter case the surface is obtained by a combination
of polygons and segments.

\bigskip
\@{\bf {Appendix B: The intersection between a sphere and a
tetrahedron}}
\smallskip
For the implementation of the Delaunay method we need to calculate the
volume of the intersection of a sphere and a tetrahedron.
In the following we use (A,B,C,D) to indicate the four points that
define the tetrahedron. The letters I, J, K and L represent any
arbitrary order of these four points.

\topinsert
\vbox{\noindent\eightpoint
{\bf Figure 13.}
Sketch of the calculation of the volume $V_{A_{I,J}}$}
\endinsert

The volume of the intersection is calculated by a sequence of
complementary volume calculations. To be specific, the intersection
volume follows by taking the volume of the whole sphere as a start.
{}From that volume we then extract the volumes cut out by each of the
planes defined by three points (I,J,K). In total there are four
of such planes, the possible permutations of (A,B,C,D). Subsequently,
we have to correct by adding each of the volumes contained in the six
spherical segments defined by two planes (I,J,K) and (I,J,L). Finally,
we should subtract the volumes of the four cones defined by the three
planes containing either I, J, K, or L. In short,
$$\eqalign{&V_{\rm intersection}=\cr
&\qquad V_{\rm sphere}-\sum_{\rm perm.} V_{P_{I,J,K}}+\sum_{\rm
perm.} V_{A_{I,J}}-
\sum_{\rm perm.} V_{S_I}\,,\cr}\eqno{\rm (B1)}$$
\@where the summations are made over the
possible permutations for I,J,K,L, and
\item{$\bullet$}
$V_{\rm sphere}$ is the volume of the sphere;
\item{$\bullet$}
$V_{P_{I,J,K}}$ is the volume of the sphere segment delineated by the
plane (I,J,K) on the side opposite to the point $L$.
\item{$\bullet$}
$V_{A_{I,J}}$ is the volume of intersection of the sphere segments
carved out by the planes (I,J,K), opposite to the point $L$ and (I,J,L),
opposite to the point $L$;
\item{$\bullet$}
$V_{S_I}$ is the volume of the intersection of the sphere segments
defined by the planes (I,K,L), opposite to $J$, (I,J,L), opposite to
$K$ and (I,J,K), opposite to $L$.

\topinsert
\vbox{\noindent\eightpoint
{\bf Figure 14.}
Sketch of the calculation of the volume $V_{S_{I}}$}
\endinsert

The calculation of $V_P$ is quite straightforward. It is given by the
expression,
$$V_P=\pi\,R^3\,(2/3-x-x^3/3),\eqno{\rm (B2)}$$
\@where $x$ is the distance of the
center of the sphere to the plane in units of the radius, $R$, or by
its complementary part in the sphere.

The calculation of $V_{A_{I,J}}$ is intrinsically more complicated.
The geometrical problem is illustrated in Figure 13 in the plane
orthogonal to the planes (I,J,K) and (I,J,L). The edge (I,J) is
indicated by the point $I$. The distance
$x$ is the distance of the centre of the sphere to this line
(expressed in units of the radius). The volume to be calculated,
$V_A$, is indicated by the doubly shaded area. It depends on $x$ and the
angle $\theta$, and for $x\tan\theta<1$ is given by

$$\eqalign{
&V_A=\cr
&\ {1 \over 3} \,\bigl\{ - {d\, x^2\, \sin\theta\,cos\theta}\cr
&\ +
{x\,(3+(3-x^2)\,\tan^2\theta)\,\sin\theta\,\cos^2\theta}
\arctan\left(-{x\,\cos\theta \over d}\right)\cr
&\ +
\arctan\left[-{x-d^2\tan\theta\over d(1-x \tan \theta)}\right]\cr
&\ -\arctan\left[{x+d^2\tan\theta\over d(1+x \tan \theta)}\right]\cr
&\ +
x\,\pi\, (3\sin^3\theta+3\cos^2\theta\,\sin\theta-x^2\,\sin^3\theta)/2
\,\bigl\} \cr}
\eqno(\rm B3)$$
\@and by the same expression plus $\pi/3$ otherwise. In this
expression, the parameter $d$ is defined by
$$d=(1-x^2)^{1/2}.\eqno(\rm B4)$$
\@This expression is strictly valid only when $0<\theta<\pi$. Otherwise
eqn. (B2) and (B3) have to be combined to get the proper answer.

The calculation of the volumes $V_S$ is even more cumbersome for its
practical implementation, but is given by a combination of equations
(B2) and (B3).
The typical geometrical situation is presented in Figure 14. By
default we assume the tip I to be located within the sphere. If this
is not the case the situation is simpler and can be reduced
quite straightforwardly to previously discussed situations and
equations.

The volume $V_{S_I}$ can also be calculated by the determination of
a sequence of complementary volumes,
$$V_{S_I}=V_{\rm tetra.}-\sum_{\rm perm.}V_{P'_I}+\sum_{\rm perm.}
V_{A_{J',K'}},\eqno{\rm (B5)}$$
\@where
$V_{\rm tetra.}$ is the volume of the tetrahedron
defined by I and the intersection points J',K',L'
of the three half-lines [I,J), [I,K) and [I,L) with the spheres.
The volume $V_{P'_I}$ is the one of the fraction
of the sphere above the plane (J',K',L') (obtained with eqn. B2),
and $V_{A_{J',K'}}$ are the three volumes of the fractions
of the sphere that are bounded by (J',K',L') and respectively
(I,J',K'), (I,J',L'), (I,K',L'), and that do not contain
I nor respectively L', K', J'. These volumes are given by a proper
use of expression (B3).

\bye